\documentclass[prd,aps,floats,12pt,nofootinbib]{revtex4}
\usepackage{times,graphicx,amssymb, amsmath, natbib}
\usepackage[T1]{fontenc}

\usepackage{epsfig}

\newcommand{\beq}{\begin{equation}}
\newcommand{\eeq}{\end{equation}}

\newcommand{\ber}{\begin{eqnarray}}
\newcommand{\eer}{\end{eqnarray}}

\def\beq{\begin{equation}}
\def\eeq{\end{equation}}

\def\ber{\begin{eqnarray}}
\def\eer{\end{eqnarray}}

\begin{document}

\title{Imprint of thawing scalar fields on large scale galaxy overdensity}

\author{Bikash R. DInda}
\email{bikash@ctp-jamia.res.in} 
\affiliation{Centre for Theoretical Physics, Jamia Millia Islamia, New Delhi-110025, India}

\author{Anjan A Sen}
\email{aasen@jmi.ac.in} 
\affiliation{Centre for Theoretical Physics, Jamia Millia Islamia, New Delhi-110025, India}

\begin{abstract}
We investigate the observed galaxy power spectrum for the thawing class of scalar field models taking into account various general relativistic corrections that occur on very large scales. We consider the full general relativistic perturbation equations for the matter as well as the dark energy fluid. We form a single autonomous system of equations containing  both the background and perturbed equations of motion which we subsequently solve for different scalar field potentials. First we study the percentage deviation from $\Lambda$CDM model for different cosmological parameters as well as in the observed galaxy power spectra on different scales in scalar field models for various choices of scalar field potentials.  Interestingly the difference in background expansion results enhancement of power from $\Lambda$CDM on small scales whereas the inclusion of GR corrections results the suppression of power from $\Lambda$CDM on large scales. This can be useful to distinguish scalar field models from $\Lambda$CDM with future optical/radio surveys. We also compare the observed galaxy power spectra for tracking and thawing types of scalar field using some particular choices for the scalar field potentials. We show that thawing and tracking models can have large differences in observed galaxy power spectra on large scales and for smaller redshifts due to different GR effects. But on smaller scales and for larger redshifts, the difference is small and is mainly due to difference in background expansion.
\end{abstract}

\maketitle
\date{\today}

\section{Introduction}
Since the first observational evidence of  the late time acceleration of the universe \citep{Perlmutter:1996ds,Riess:1998cb}, it has been the biggest challenge in theoretical as well as observational cosmology to find the source of the repulsive gravitational force that causes this acceleration. We still do not have confirmatory evidence whether this is due to some extra dark component in the energy budget of the universe (commonly known as "dark energy") \citep{2000IJMPD...9..373S,2003RvMP...75..559P,2003PhR...380..235P} or due to some modification of Einstein gravity on cosmological scales \citep{Tsujikawa:2010zza}. Although the Planck-2015 result \citep{Ade:2015xua} shows that the concordance $\Lambda$CDM universe (containing Cosmological Constant $\Lambda$ and Cold Dark Matter) is consistent with a whole set of observational data, still there are some recent observational results that indicate inconsistency with $\Lambda$CDM model \citep{2014ApJ...793L..40S,2015A&A...574A..59D,Riess:2016jrr,Bonvin:2016crt}. This is in addition to the theoretical inconsistencies  like fine tuning\citep{2000IJMPD...9..373S} and cosmic coincidence problems\citep{2000IJMPD...9..373S} that are present in $\Lambda$CDM model. This motivates people to go beyond $\Lambda$CDM paradigm and consider models with evolving dark energy. 

But one needs to distinguish this evolving dark energy from a cosmological constant ($\Lambda$) which does not change through out the history of the universe. To do so, we need to study the effect of dark energy on different cosmological observables related to background expanding universe as well as to the process of growth of structures in our universe. This can be done through observations with Supernova Type-Ia as standard candles \citep{Betoule:2014frx}, or observing the fluctuations in the temperature of cosmic microwave background radiation (CMBR) \citep{Ade:2015xua} or looking at the galaxies and their distribution over different distance scales as well as at different redshifts \citep{Alam:2016hwk}. The latest Planck results in 2015 together with data from Supernova Type-Ia and also data related to Baryon Acoustic Oscillations (BAO) from different redshift surveys have put unprecedented constraints on different cosmological parameters including those related to dark energy properties \citep{Ade:2015xua,Ade:2015rim}. But as far as the dark energy is concerned, we have mainly constrained its background evolution till now. This is because most of the observations mentioned above are either related to background universe, or perturbed universe on sub-horizon scales where Newtonian treatment is sufficient and one can safely ignore the fluctuations in dark energy. Hence it has not been possible till now to probe the inhomogeneity in dark energy which can be very useful in distinguishing different dark energy models.

Future optical as well as radio/infrared surveys like LSST \citep{Abell:2009aa}, SKA \citep{Maartens:2015mra} etc, have potential to cover larger sky area and to extend to much higher redshifts probing horizon scales and beyond. This will give us a whole lot of information about our universe on such large length scales which is not known yet. Crucially on these scales, one needs to consider the full general relativistic (GR) treatment to study how fluctuations grow and dark energy perturbation can not be neglected any more. This can be a smoking gun to distinguish evolving dark energy model from $\Lambda$CDM as $\Lambda$, being a constant is not perturbed whereas any other evolving dark energy component should be perturbed and hence affects the growth of matter fluctuations  on large scales in a different way than the $\Lambda$CDM model.

Apart from this extra effect coming from the dark energy perturbations on large scales, there are other GR effects on galaxy overdensity on large scales \citep{Yoo:2009au,Bonvin:2011bg,Bonvin:2014owa,Challinor:2011bk,Jeong:2011as,Yoo:2012se,Bertacca:2012tp,Duniya:2016ibg,Duniya:2015dpa}. Primarily there are two sources of GR effects on large scales. One is related to the  gravitational potential \citep{Duniya:2016ibg}  which can be local or at the observed galaxies or along the line of sight and the other is related to the peculiar velocities due to the motion of galaxies relative to the observer. In recent years, there has been number of studies related to the calculation of power spectrum of galaxy overdensity on large scales taking into account the dark energy perturbation as well as various GR effects on large scales. This has been done mostly in the context of $\Lambda$CDM universe \citep{Duniya:2016ibg}.

The simple way to consider evolving dark energy model is a canonical scalar field rolling over its potential. The first such model considered was the "tracker" scalar field model \citep{Wetterich:1987fm,2006IJMPD..15.1753C,1988PhRvD..37.3406R,1998PhRvL..80.1582C,1999PhRvD..59b3509L,1999PhRvD..59l3504S} with some particular types of potential that causes the scalar field to track the background radiation/matter component until recent past when slope of the potential changes so that it can behave as $\Lambda$ and causes the universe to accelerate. This tracker behaviour helps to evade the cosmic coincidence problem that is present in the $\Lambda$CDM model. The full observed galaxy power spectrum that incorporates various GR corrections on horizon scales  has been studied recently by Duniya et al \citep{Duniya:2013eta} for tracker scalar field models.

There is another kind of scalar field models, ``the thawer class'', \citep{Caldwell:2005tm} where the scalar field is initially frozen at some flat part of the potential due to large Hubble friction in the early time and behaves like a cosmological constant with $w \approx -1$. Later on, as the Hubble friction decreases, the scalar field slowly thaws away from its initial frozen state and evolves away from cosmological constant type behaviour. In this case, the scalar field never evolves much from its initial frozen state and the equation of state always remain very close to $w=-1$. Thawer scalar fields are much similar to the inflaton that drives the acceleration in the universe in the early time. It is also interesting to note that the thawer canonical scalar field model has a generic analytical behaviour for its equation of state for nearly flat potentials \citep{Scherrer:2007pu} ( Also see \citep{Li:2016grl} for the generalisation of this result for non-canonical scalar field). Thawing model in the context of tachyon field \citep{Ali:2009mr} as well as galileon field \citep{Hossain:2012qm} have been also considered in the recent past.To best of our knowledge, there has been no study till date on observed galaxy power spectrum for thawer class of models that incorporates various GR corrections on large scales.

In this paper we study the full general relativistic treatment for the growth of linear fluctuations in cosmological models with thawing scalar field as dark energy. We form a single set of autonomous system of eight coupled equations involving both the background as well as the perturbed equations of motion and solve it for various scalar field potentials. Subsequently we study the power spectrum for the observed  galaxy overdensity taking into account various GR corrections for different scalar field potentials and compare them with $\Lambda$CDM model. We also study the difference in observed galaxy power spectrum for thawing and tracking/freezing class of models for some specific choices of potential.

The paper is organised as follows: in section 2, we briefly describe the background equations for the thawing dark energy models; in section 3, we describe the full general relativistic perturbation equation for linear fluctuations in both dark energy and matter, form a single set of autonomous equations involving both background evolution and evolution for the fluctuations and study various cosmological quantities; in section 4, we calculate the observed galaxy power spectrum taking into account various GR correction terms for different scalar field potentials  and compare them with $\Lambda$CDM model; in section 5, we study the difference between thawing and tracking class models; finally in section 6, we write our conclusions.

\section{Background evolution}

We consider flat FLRW background geometry for our Universe with $a(t)$ being the scale factor. The Lagrangian density for a minimally coupled canonical scalar field  $\phi$ is given by

 \noindent
\begin{equation}
\mathcal{L} = \dfrac{1}{2} (\partial^{\mu}\phi) (\partial_{\mu}\phi) - V(\phi).
\end{equation}

\noindent
Here $ V(\phi) $ is the potential for the field $\phi$. The background energy density and pressure of the scalar field are given by 

\begin{eqnarray}
\bar{\rho_{\phi}} = \frac{1}{2} \dot{\phi}^{2} + V(\phi), \nonumber\\
\bar{P_{\phi}} = \frac{1}{2} \dot{\phi}^{2} - V(\phi),
\end{eqnarray} 

\noindent
where overdot represents derivative with respect to the cosmic time $ t $. The equation of motion for the scalar field is given by

\begin{equation}
\ddot{\phi} + 3 H \dot{\phi} + V_{\phi} = 0,
\end{equation}

\noindent
where $ H $ is the Hubble parameter and subscript $ \phi $ is the derivative with respect to the field $ \phi $. The background FRW equation is given by 

\begin{equation}
3 M_{pl}^2 H^{2} = \bar{\rho}_{m} + \bar{\rho}_{\phi},
\end{equation}

\noindent
where $M_{pl} = (8\pi G)^{-1/2}$, $G$ being the Newton's gravitational constant, is the Planck's mass. $\bar{\rho}_{m}$ is the energy density of the background matter component which includes contribution from dark matter and baryons.

\section{Relativistic perturbations with scalar field}

We assume conformal Newtonian gauge with vanishing anisotropic stress for the perturbed space-time:

\begin{equation}
ds^{2} = a^{2}(\tau)[(1+2\Phi) d\tau^{2}-(1-2\Phi) d\vec{x}.d\vec{x}],
\end{equation}

\noindent
where $ \tau $ is the conformal time, $ \Phi $ is the gravitational potential and $\vec{x}$ are the comoving coordinates. The linearised Einstein equations are now given by \citep{2008PhRvD..78l3504U}:

\begin{equation}
\nabla^{2} \Phi - 3 \mathcal{H} (\Phi' + \mathcal{H} \Phi) = 4 \pi G a^{2} \sum_{i} \delta \rho_{i},
\end{equation}

\begin{equation}
\Phi' + \mathcal{H} \Phi = 4 \pi G a^{2} \sum_{i} (\bar{\rho_{i}} + \bar{P_{i}}) v_{i},
\end{equation}

\begin{equation}
\Phi'' + 3 \mathcal{H} \Phi' + (2 \mathcal{H}' + \mathcal{H}^{2}) \Phi = 4 \pi G a^{2} \sum_{i} \delta P_{i},
\end{equation}

\noindent
where prime is the derivative with respect to the conformal time $ \tau $, $ \mathcal{H} $ is the conformal Hubble parameter, $ \bar{\rho_{i}} $ and $ \bar{P_{i}} $ are the background energy density and pressure for the individual fluid $ i $ (here $i$ stands for either `m' for matter or `$\phi$' for scalar field) and $ \delta \rho_{i} $, $ \delta P_{i} $ and $ v_{i} $ are the perturbation to individual component's background energy density, pressure and velocity field respectively. The irrotational part of individual velocity field is given by $ \vec{v_{i}} = - \vec{\nabla} v_{i} $. Combining eqns. (6) and (7) we can get the relativistic Poisson equation as 

\begin{equation}
\nabla^{2} \Phi = 4 \pi G a^{2} \sum_{i} \bar{\rho_{i}} \Delta_{i},
\end{equation}

\noindent
where $ \Delta_{i} = \delta_{i} + 3 \mathcal{H} (1 + w_{i}) v_{i} $ is the comoving energy density contrast for individual component and they are the correct tracer of the gravitational potential on large scales.
\\
From the conservation of stress-energy tensor we can get relativistic continuity and Euler equations as

\begin{equation}
\delta' + 3 \mathcal{H} (\dfrac{\delta P}{\delta \rho} - \dfrac{\bar{P}}{\bar{\rho}}) \delta = (1 + \dfrac{\bar{P}}{\bar{\rho}}) (\theta + 3 \Phi')
\end{equation}

\noindent
and

\begin{equation}
\theta' + 3 \mathcal{H} (\dfrac{1}{3} - \dfrac{\bar{P}'}{\bar{\rho}'}) \theta = \dfrac{\nabla^{2} \delta P}{\bar{\rho} + \bar{P}} + \nabla^{2} \Phi
\end{equation}

\noindent
respectively where $ \theta = - \vec{\nabla}.\vec{v} $ and $ \delta $ is defined as $ \delta \rho = \bar{\rho} \delta $. Note that the above continuity and Euler equations are valid for individual components e.g. matter or scalar field and also for the combined (matter plus scalar field) one.

For scalar field $\phi$, the perturbed energy density, pressure and velocity (at linear order) are given by 

\begin{equation}
\delta \rho_{\phi} = \frac{{\phi'} {(\delta \phi')}}{a^2} - \frac{{\phi'}^{2} \Phi}{a^2} + V_{\phi} \delta \phi,
\end{equation}

\begin{equation}
\delta P_{\phi} = \frac{{\phi'} {(\delta \phi')}}{a^2} - \frac{{\phi'}^{2} \Phi}{a^2} - V_{\phi} \delta \phi,
\end{equation}

\noindent
and

\begin{equation}
a (\bar{\rho_{\phi}} + \bar{P_{\phi}}) v_{\phi} = \frac{{\phi'}}{a} (\delta \phi),
\end{equation}

\noindent
where $ \delta \phi $ is the perturbation to the background field $ \phi $. 

We now define the following dimensionless quantities related to the background \citep{Scherrer:2007pu} as well as perturbed universe:

\begin{eqnarray}
x &=& \frac{\Big{(} \dfrac{d \phi}{d N} \Big{)}}{\sqrt{6} M_{Pl}}, \hspace{1 cm}  y = \frac{\sqrt{V}}{\sqrt{3} H M_{Pl}}, \nonumber\\
\lambda &=& - M_{Pl} \frac{V_{\phi}}{V}, \hspace{1 cm}  \Gamma = V \frac{V_{\phi \phi}}{V_{\phi}^{2}}, \nonumber\\
\Omega_{\phi} &=& x^{2} + y^{2}, \hspace{1 cm} \gamma = 1 + w_{\phi} = \frac{2 x^{2}}{x^{2} + y^{2}},\nonumber\\
q &=&  (\delta \phi)/\dfrac{d \phi}{d N}.
\end{eqnarray} 

\noindent
where $N=log(a)$ is the number of e-folding. The $x$ here is different from the comoving coordinates in equation (5). Here $\Omega_{\phi}$ is the density parameter related to the scalar field, $w_{\phi}$ is the equation of state for the scalar field. With these quantities, one can now form a single set of autonomous system of equations involving quantities related to both the background \citep{Scherrer:2007pu} as well as perturbed universe as following:

\begin{eqnarray}
\dfrac{d \gamma}{d N} &=& 3 \gamma (\gamma - 2) + \sqrt{3 \gamma \Omega_{\phi}} (2 - \gamma) \lambda, \nonumber\\
\dfrac{d \Omega_{\phi}}{d N} &=& 3 (1 - \gamma) \Omega_{\phi} (1 - \Omega_{\phi}), \nonumber\\
\dfrac{d \lambda}{d N} &=& \sqrt{3 \gamma \Omega_{\phi}} \lambda^{2} (1 - \Gamma),\nonumber\\
\dfrac{d \mathcal{H}}{d N} &=& -\frac{1}{2} (1 + 3 (\gamma-1) \Omega_{\phi}) \mathcal{H}\nonumber\\
\dfrac{d \Phi}{d N} &=& \Phi_{1}\nonumber\\
\dfrac{d q}{d N} &=& q_{1}\nonumber\\
\dfrac{d \Phi_{1}}{d N} &=& -(1 + B) \Phi_{1} - (2 B - 3 + 1.5 \gamma \Omega_{\phi}) \Phi + 1.5 \gamma\Omega_{\phi} \Big{[} q_{1}+ (2 g - B) q \Big{]}\nonumber\\
\dfrac{d q_{1}}{d N } &=& - (2 g - B) q_{1} - B_{q} q + 4 \Phi_{1} + 2 g \Phi.
\end{eqnarray} 

\noindent
Here $ B = 1.5(1 - (\gamma-1) \Omega_{\phi}) $, $B_{q} = 6 g - \dfrac{dB}{dN} - 2 B g + \dfrac{k^{2}}{\mathcal{H}^{2}} $ and $ g = \sqrt{1.5} \lambda y^{2}/x $ where $x$ and $y$ can be easily written in terms of $\gamma$ and $\Omega_{\phi}$ using equation (15).

\noindent
Note that in the above set of equations, equations involving $\Phi$ and $q$ are written in Fourier space where for simplicity we have taken the same notations in the Fourier space for corresponding entities.  Using Fourier version of eqns. (6) and (12) and changing the $ \tau $ derivative to the derivative w.r.t $ N $, we get matter density contrast as given by

\begin{equation}
\delta_{m} = - \dfrac{2}{\Omega_{m}} \Big{[} \dfrac{d\Phi}{dN} + \Big{(} 1 - x^{2} + \dfrac{k^{2}}{3 \mathcal{H}^2} \Big{)} \Phi + x^{2} \Big{(} \dfrac{dq}{dN} - B q \Big{)} \Big{]}.
\end{equation} 

\noindent
Similarly, using Fourier version of eqs. (7) and (14), we get peculiar velocity for matter as 

\begin{equation}
y_{m} = 3 \mathcal{H} v_{m} = \dfrac{2}{\Omega_{m}} \Big{[} \dfrac{d\Phi}{dN} + \Phi - 3 x^{2} q \Big{]}.
\end{equation}

\noindent
Using eqs. (17) and (18), we get co-moving matter density contrast as $ \Delta_{m} = \delta_{m} + y_{m} $ which is gauge invariant quantity.

\subsection{Initial conditions}     
  
To solve the system of equations (16), we need initial conditions for ($\gamma, \Omega_{\phi}, \lambda, \mathcal{H}$) for the background universe and  ($\Phi, \dfrac{d\Phi}{dN}, q, \dfrac{d q}{dN} $) for the perturbed universe. We set our initial condition at decoupling ($z=1000$), which ensures that there is negligible dark energy contribution at that time and the universe is matter dominated. 

As we are considering the thawing class of models where the scalar field is initially frozen due to large Hubble friction at $w_{\phi} \sim -1$, $\gamma_{i} \sim 0$ initially. We set it at $10^{-7}$. Our results are not very sensitive to this value as long $\gamma_{i} <<1$. The initial value for $\Omega_{\phi}$ is also negligible (at $z=1000$, we do not expect any contribution from dark energy). Initial value for $\lambda$, the slope of the potential, is an important quantity. It determines the subsequent evolution of the scalar field. For $\lambda_{in} <<1$, the scalar field does not evolve much from its initial frozen state and always stays very close to the cosmological constant behaviour. For large values of $\lambda_{i}$, the scalar field can thaws away substantially from its initial frozen state $w \sim -1$. In our case, we fix the initial values for $\Omega_{\phi}$ and $\mathcal{H}$ to set $\Omega_{\phi}$, and ${\mathcal{H}_{0}}$  at present ($z=0$) to their desired values.

For the scalar field perturbation, we assume it is negligible initially as there is hardly any contribution from dark energy at $z=1000$ and we set $q= \dfrac{d q}{dN} = 0$ initially.

To set the initial condition for the gravitational potential $\Phi$, we know that, during matter domination, $\Phi$ is constant and hence $\dfrac{d \Phi}{dN}=0$ initially. Also during matter domination, $\Delta_{m} \sim a$ and using the Poisson equation, one can easily show

\begin{equation}
\Phi_{in} = - \frac{3}{2} \frac{\mathcal{H}^{2}_{in}}{k^2} a_{in}.
\end{equation}

\begin{center}
\begin{figure*}
\begin{tabular}{c@{\quad}c}
\epsfig{file=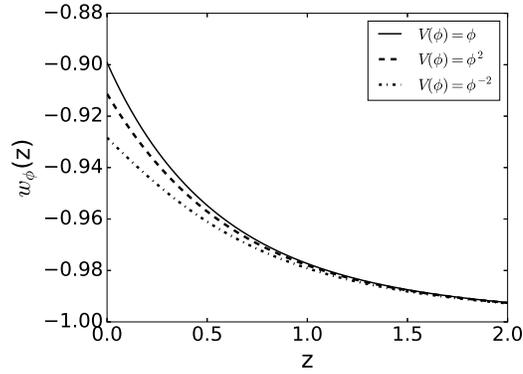,width=7.5 cm}
\end{tabular}
\caption{Behaviour of the Equation of state for the scalar field $w_{\phi}$ as a function of redshift for different potentials. $\Omega_{m0} = 0.28$ and $\lambda_{i} = 0.7$ in these plots.
}
\end{figure*}
\end{center}

\begin{center}
\begin{figure*}
\begin{tabular}{c@{\quad}c}
\epsfig{file=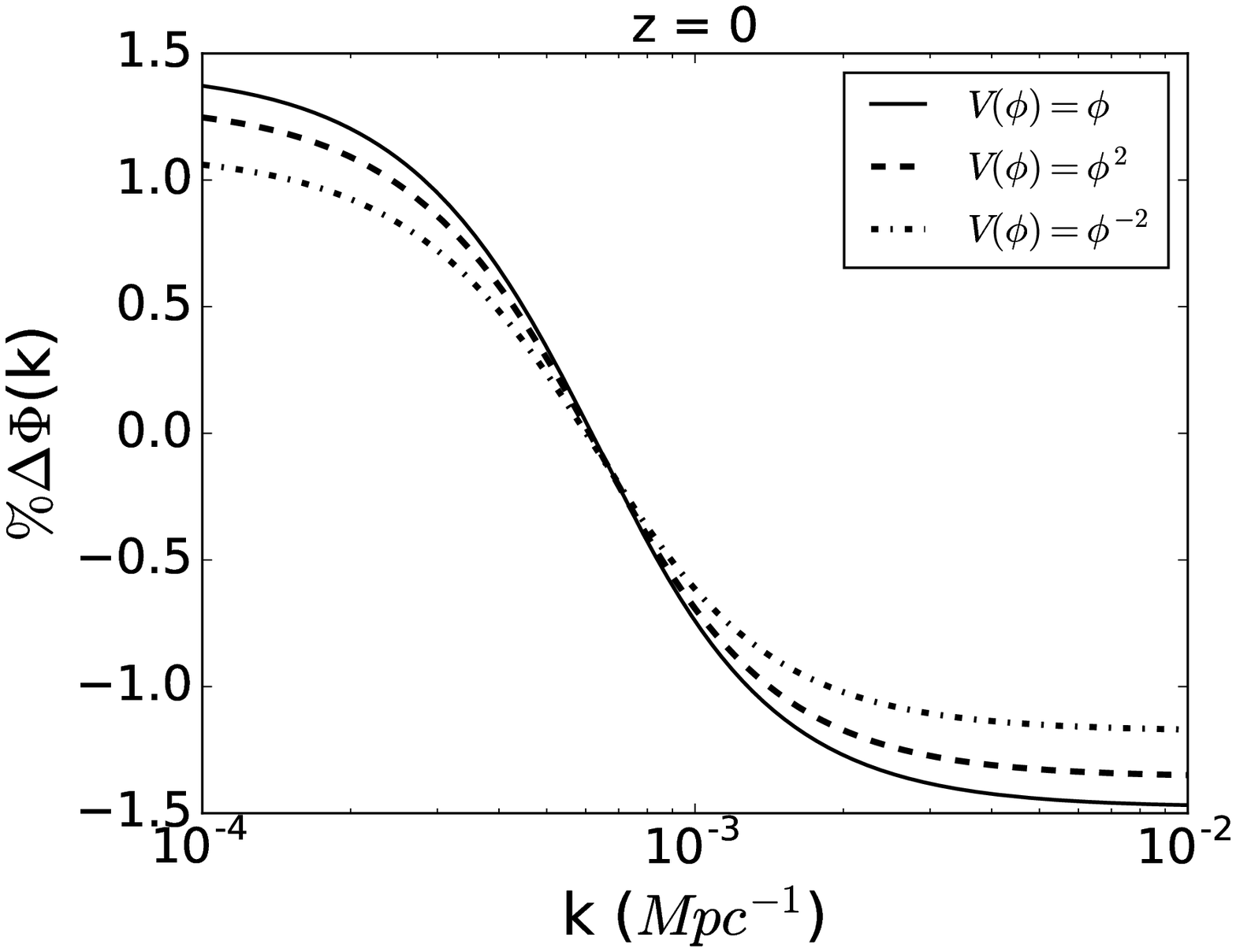,width=7.5 cm}
\epsfig{file=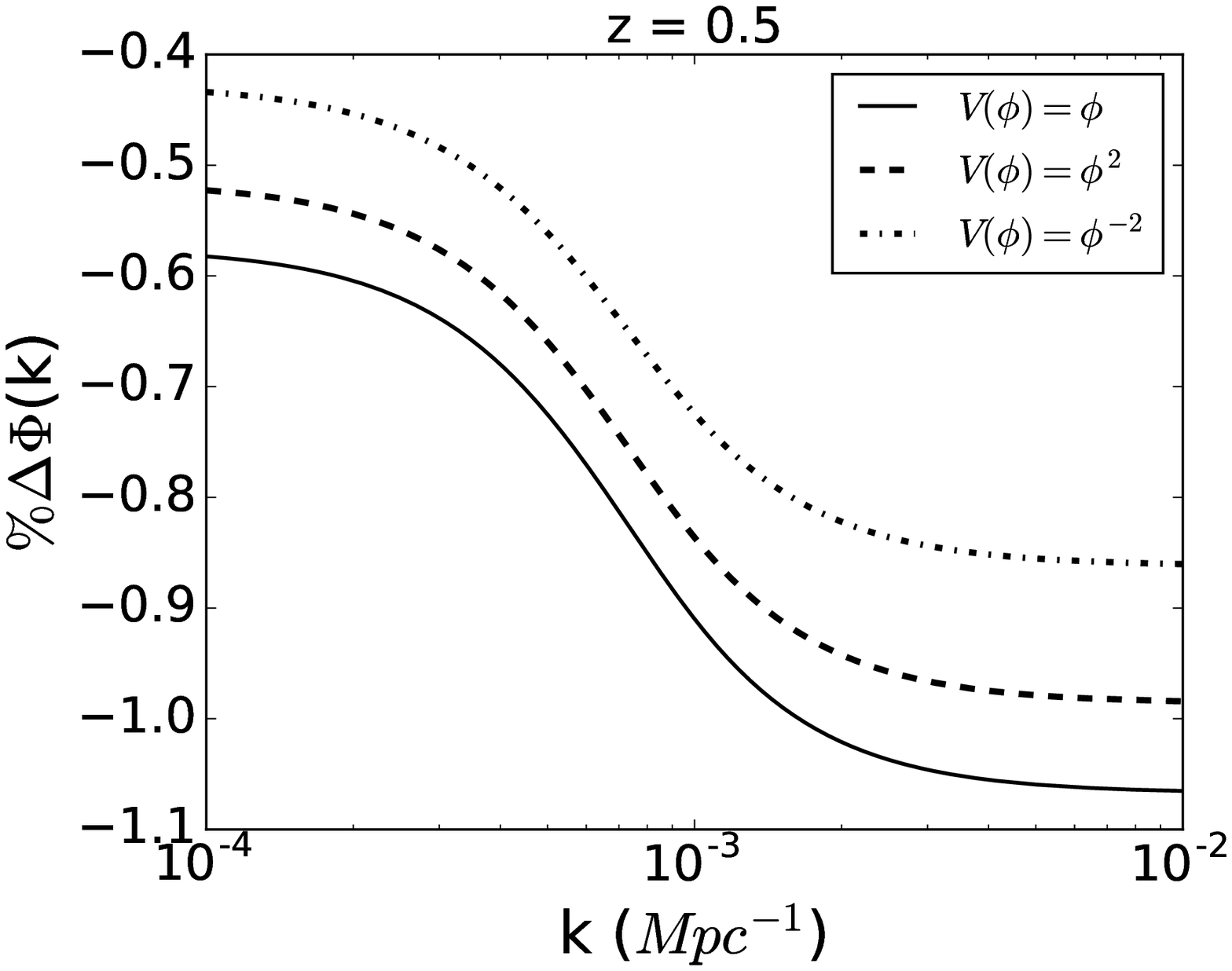,width=7.5 cm}\\
\epsfig{file=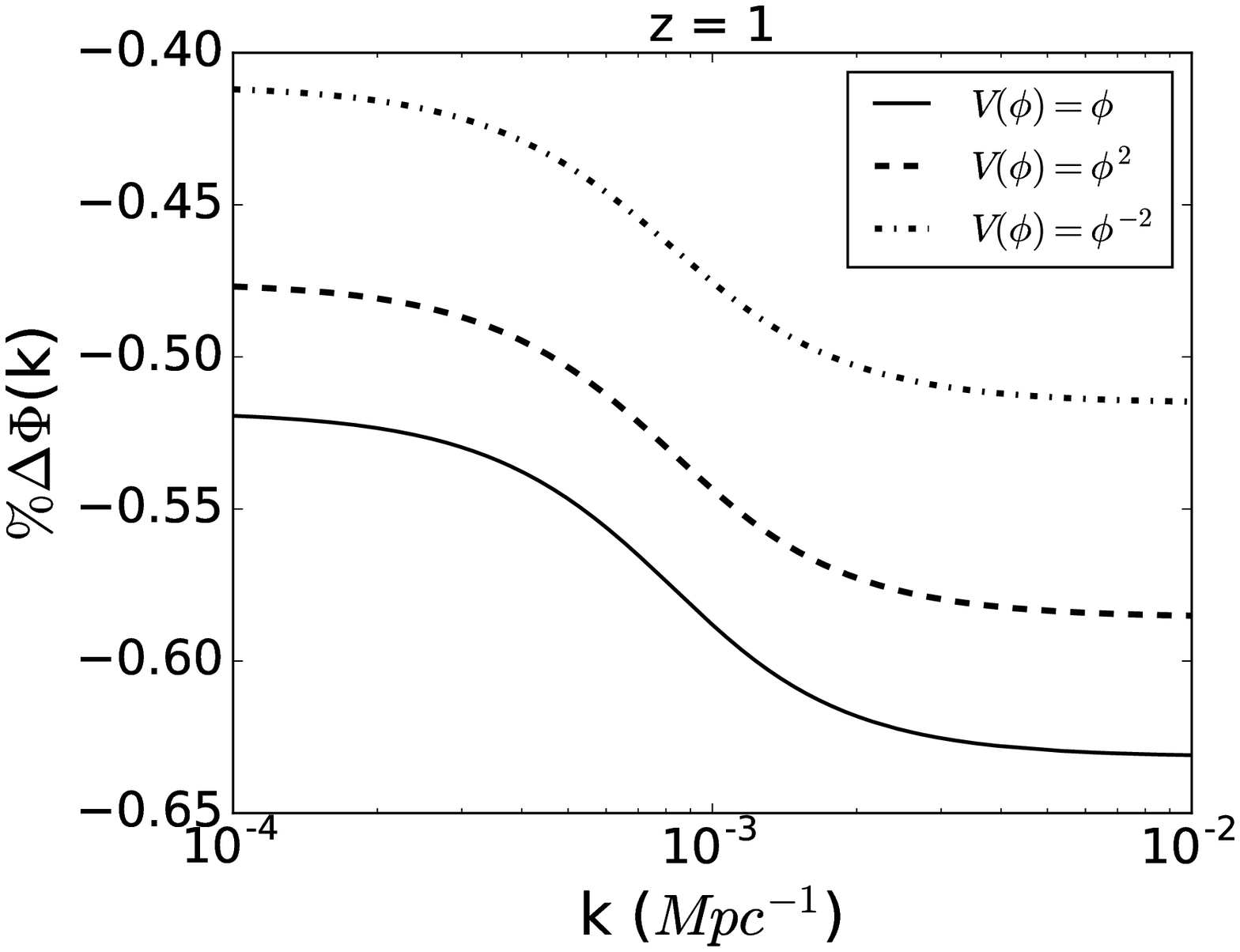,width=7.5 cm}
\epsfig{file=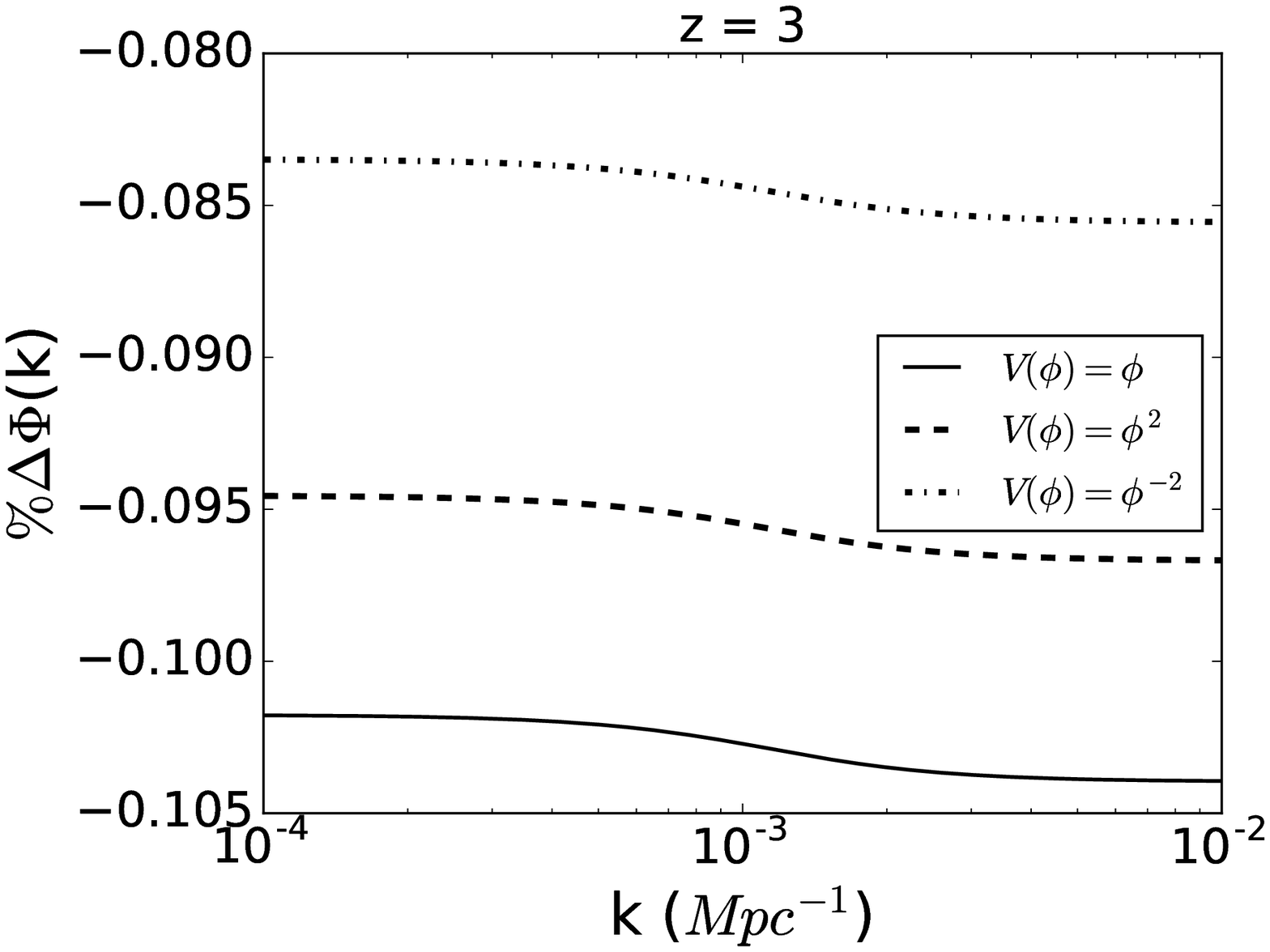,width=7.5 cm}
\end{tabular}
\caption{Percentage deviation in $ \Phi $ from $ \Lambda$CDM model: negative values
in y-axis means they are all suppressed from $ \Lambda $CDM. $\Omega_{m0} = 0.28$ and $\lambda_{i} = 0.7$ in these plots. Here and in subsequent plots, $\% \Delta X = (X^{de} / X^{\Lambda} - 1)\times 100$.
}
\end{figure*}
\end{center}

\begin{center}
\begin{figure*}
\begin{tabular}{c@{\quad}c}
\epsfig{file=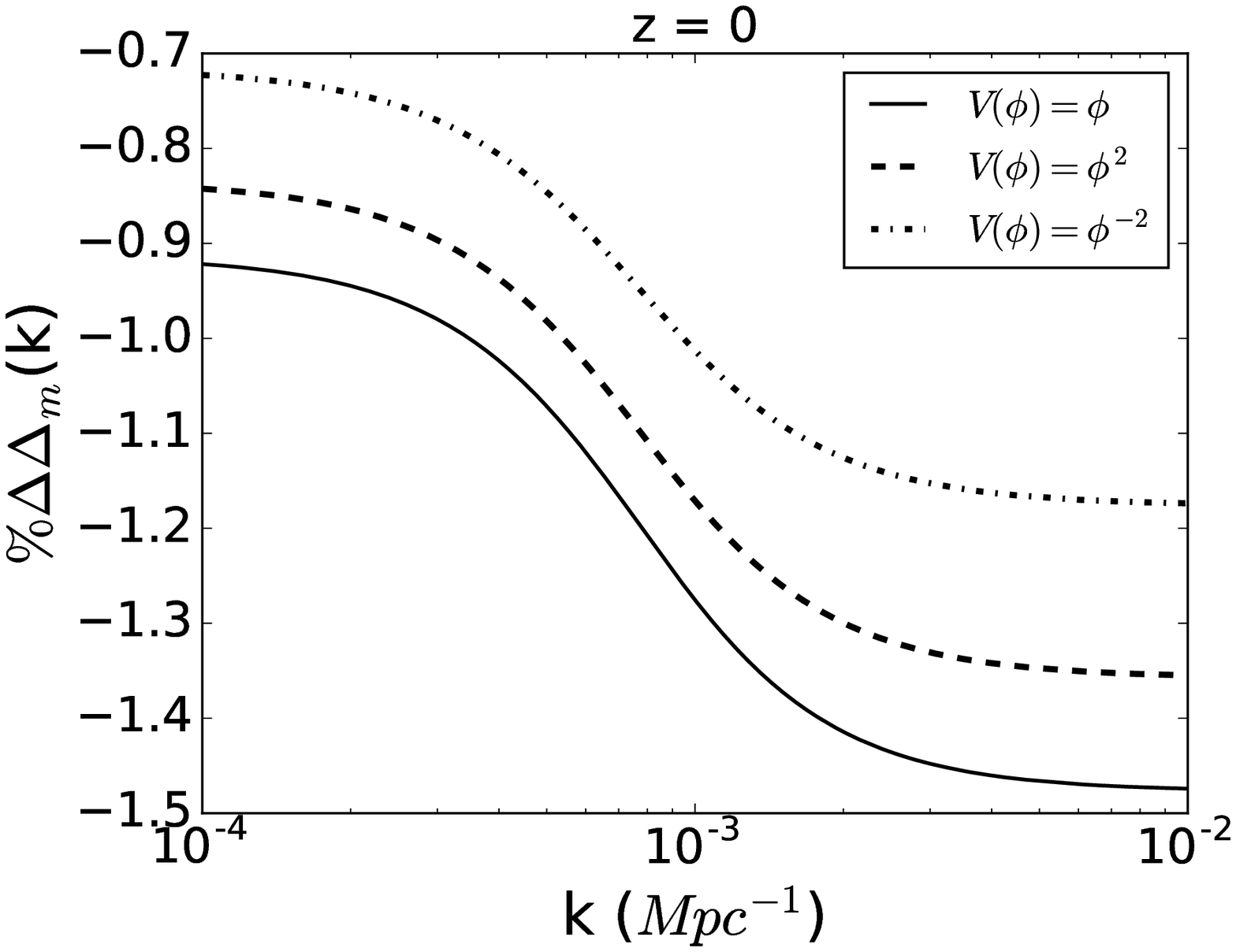,width=7.5 cm}
\epsfig{file=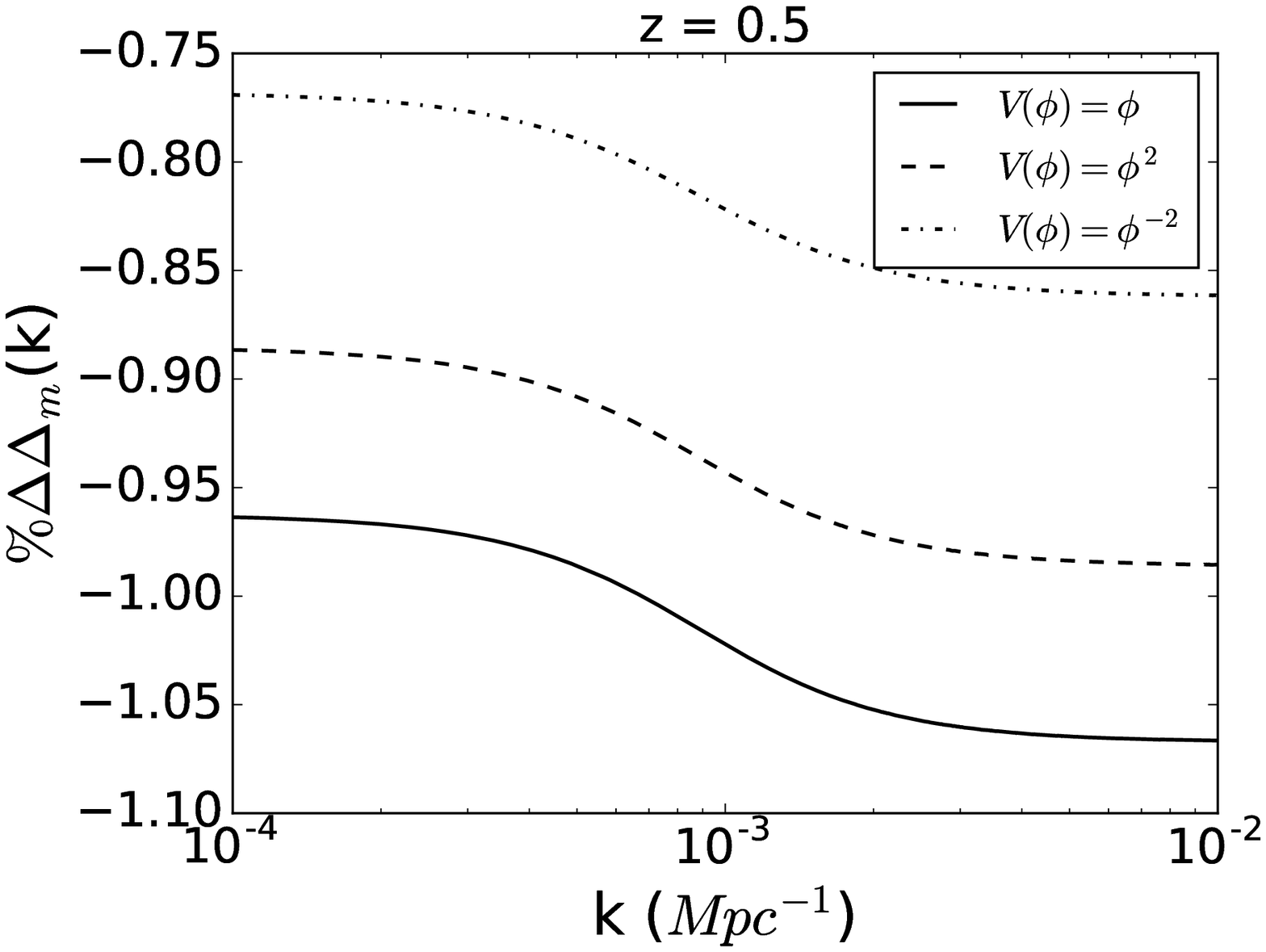,width=7.5 cm}\\
\epsfig{file=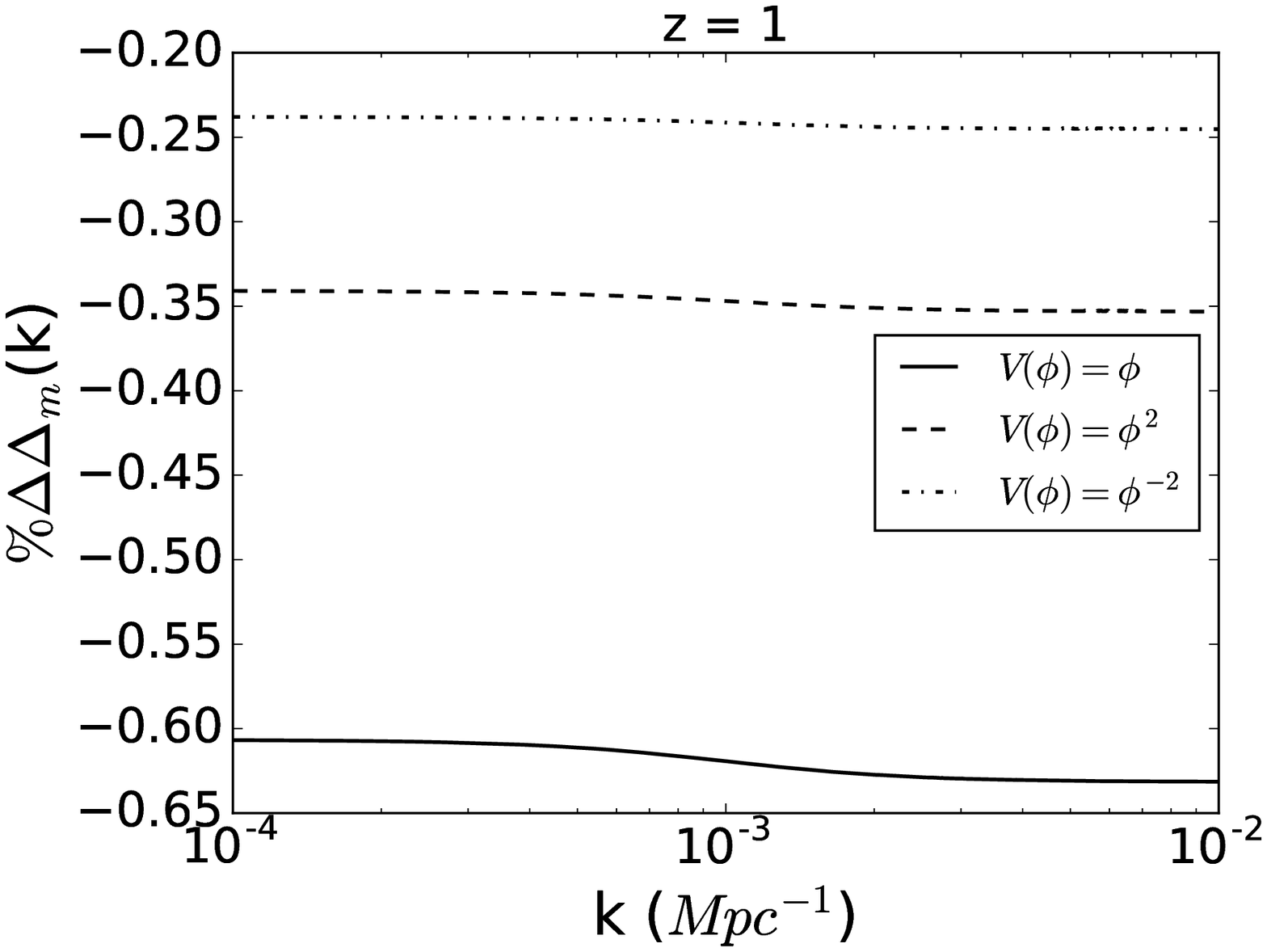,width=7.5 cm}
\epsfig{file=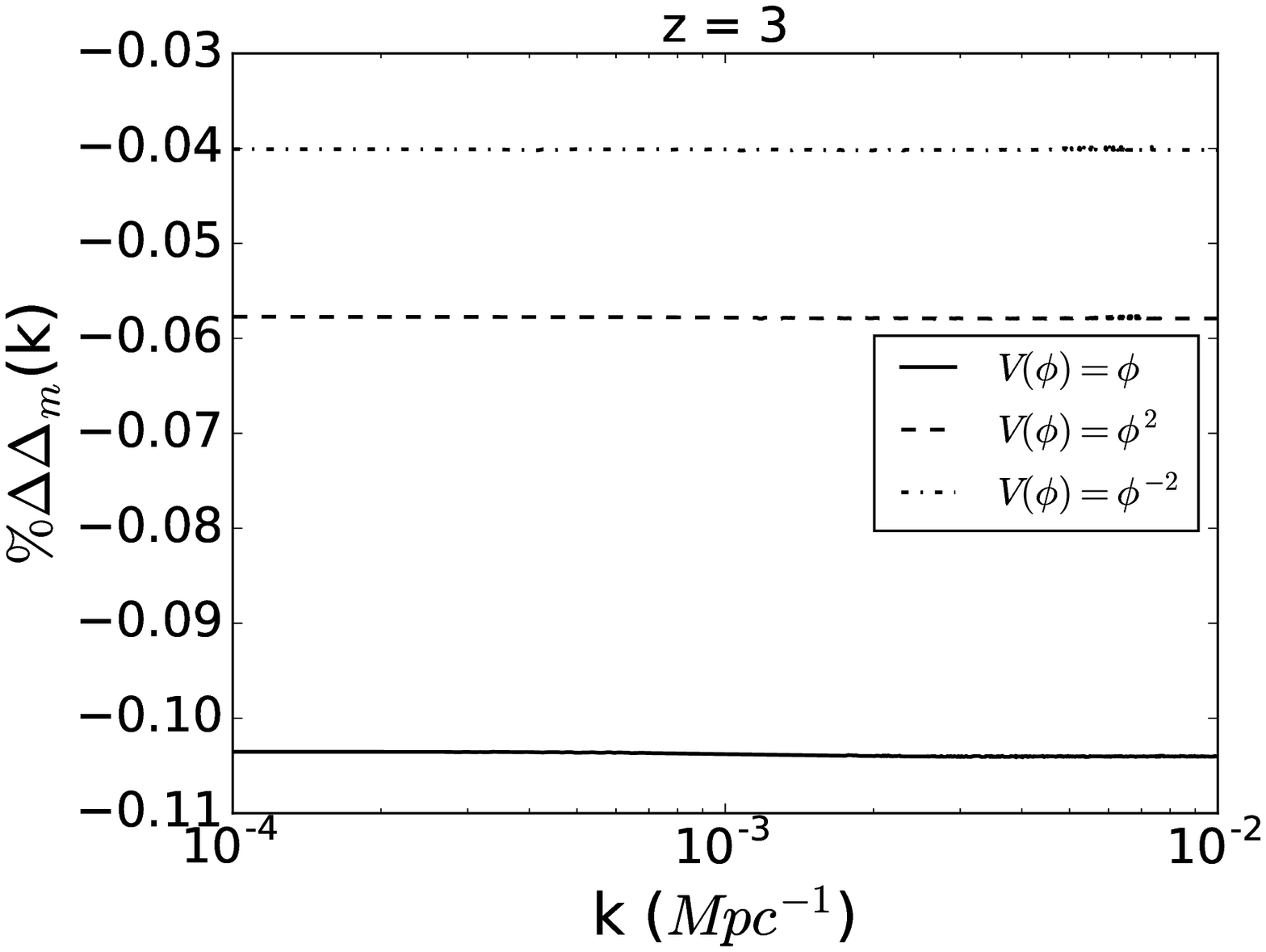,width=7.5 cm}
\end{tabular}
\caption{Percentage deviation in comoving density contrast $ \Delta_{m} $ from $ \Lambda$CDM model.
}
\end{figure*}
\end{center}
\begin{center}
\begin{figure}
\begin{tabular}{c@{\quad}c}
\epsfig{file=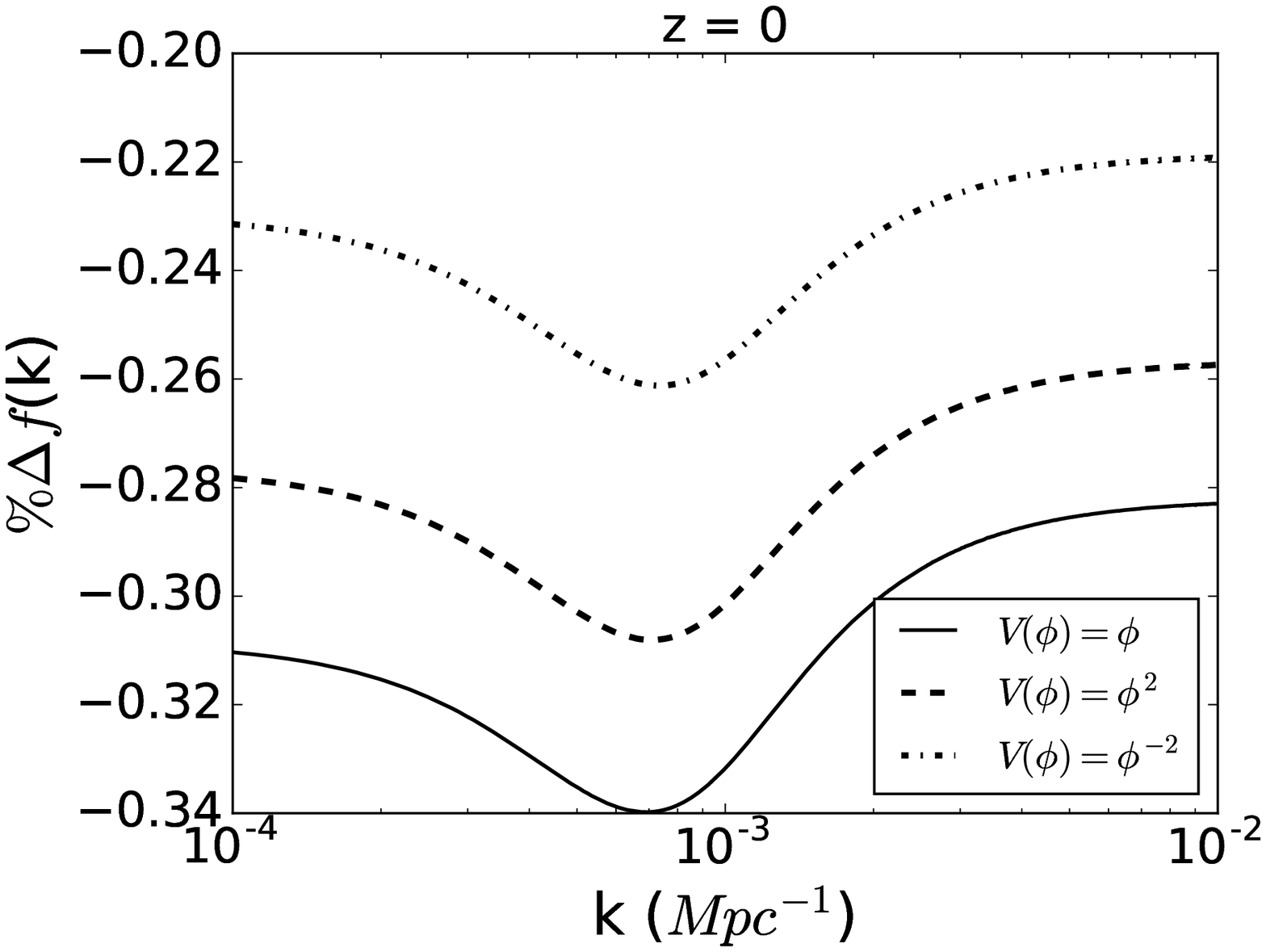,width=7.5 cm}
\epsfig{file=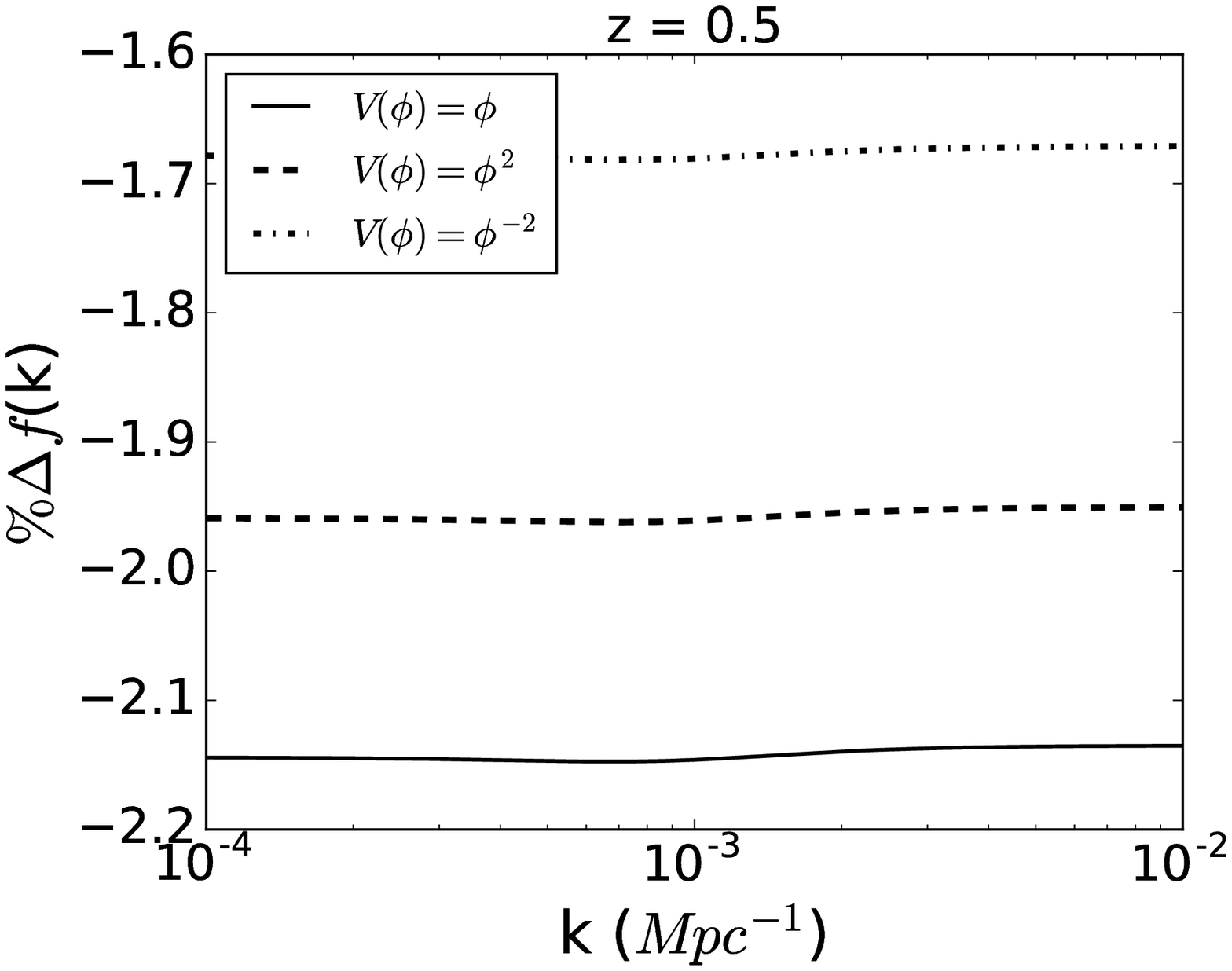,width=7.5 cm}\\
\epsfig{file=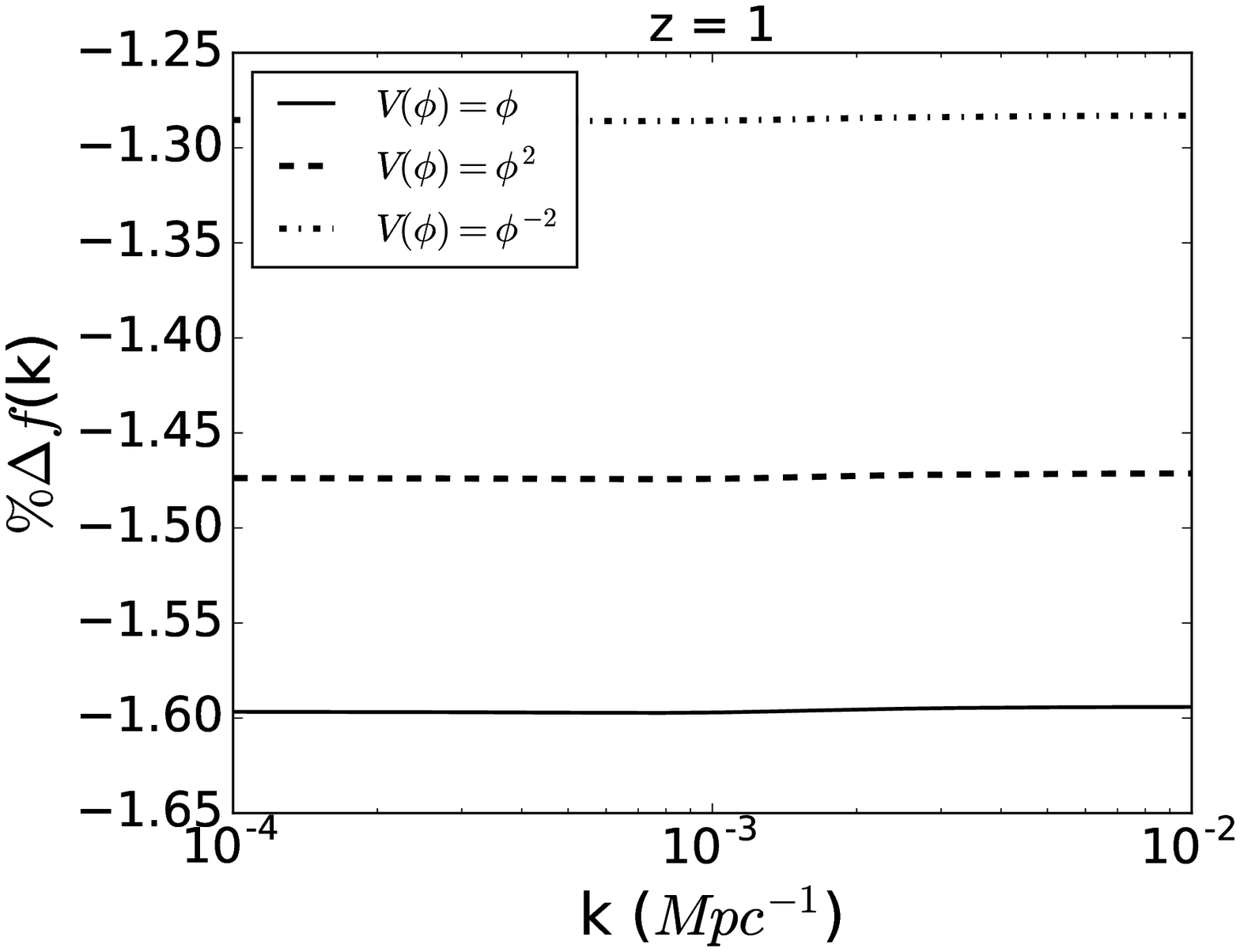,width=7.5 cm}
\epsfig{file=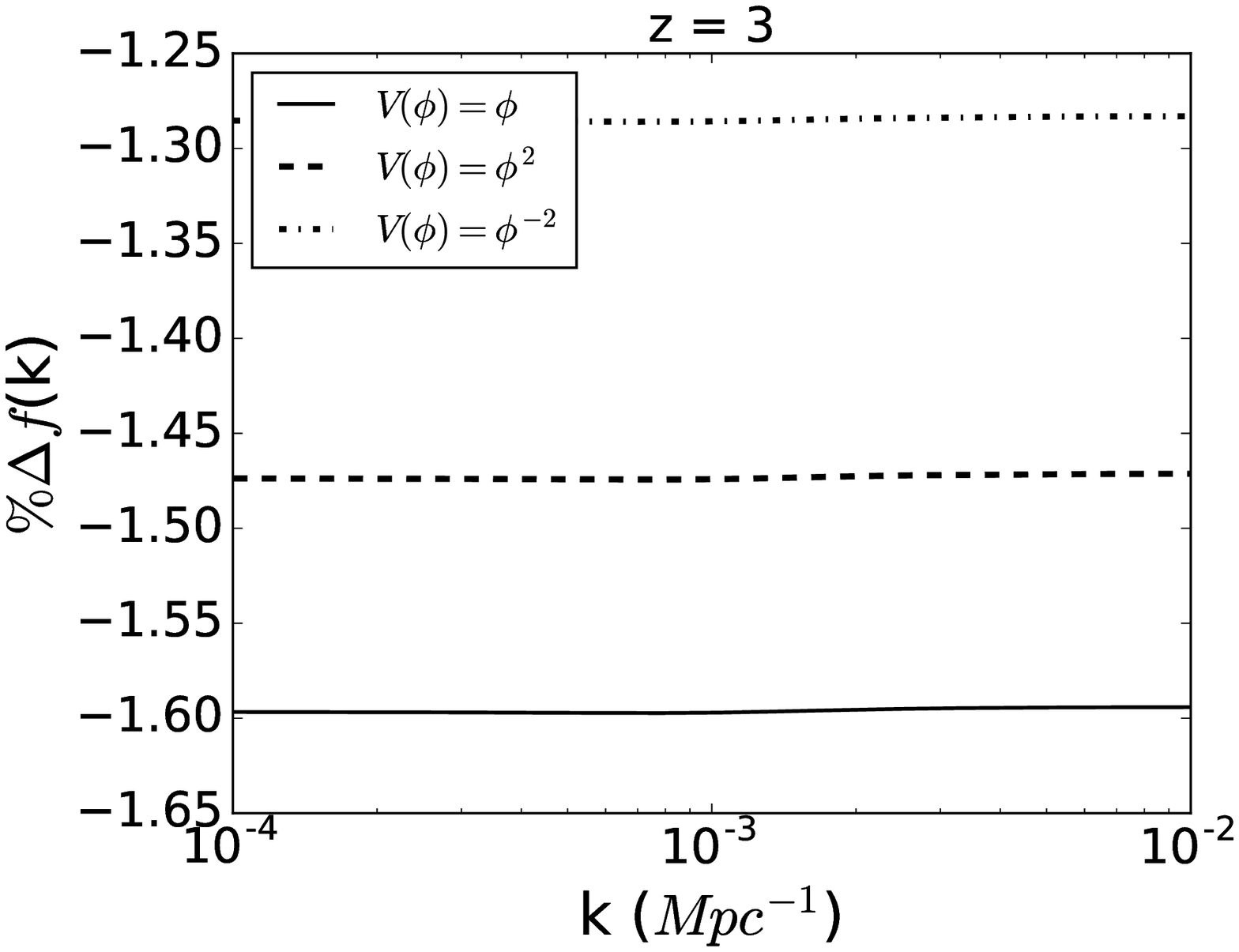,width=7.5 cm}
\end{tabular}
\caption{Percentage deviation in $f$ from $ \Lambda$CDM model.}
\end{figure}
\end{center}

\subsection{Behaviour of cosmological parameters}

With the system of autonomous equations given in (16) and the initial conditions described above, we solve the set of equations and study various cosmological parameters. For this purpose we concentrate on power-law potentials, more specifically the linear, squared and inverse-squared potentials. It is straightforward to generalise our study for more exotic potentials. We also fix the following quantities for our purpose: 
$\Omega_{m0} = 0.28$, $\lambda_{i} = 0.7$ and $H_{0} = 70 km/s/Mpc$. 

In figure 1, we show the behaviour of the equation of state as a function of redshift for different potentials. Remember the equation of state for the scalar field $w_{\phi} = \gamma -1$ where $\gamma$ is solved using the autonomous system of equations (16) for different potentials. As a thawing model, for every potential, $w_{\phi}$ starts from $-1$ at $z\sim 1000$ and  all of them are fixed at $\lambda_{i} = 0.7$. This sets the identical initial conditions for all the potentials.

Next we study behaviour of gravitational potential. In figure 2, we show  the deviation in the gravitational potential for scalar field models with different potentials from $\Lambda$CDM model. In this plot and in subsequent plots $\%\Delta X =100 (X_{de} - X_{\Lambda CDM})/ X_{\Lambda CDM}$ for any cosmological parameter $X$. For lower redshifts and smaller scales there is a suppression from the $\Lambda$CDM model which is solely due to the difference in background expansion as on smaller scales the dark energy perturbation does not contribution. On larger scales the enhancement in scalar field model is due to the extra contribution in the gravitational potential from the dark energy perturbation. For larger redshifts, the deviation decreases as the dark energy contribution diminishes with redshifts and all models including $\Lambda$CDM starts behaving like matter-only models.

In figure 3, we show the same  behaviour for the gauge invariant matter density contrast $\Delta_{m}$. As one can see, the deviation from $\Lambda$CDM is less than $1\%$ for all scales and redshifts. Also except for very low redshifts, there is hardly any scale dependence, showing that the effect of dark energy perturbation is very small and whatever deviation from $\Lambda$CDM is present, is due to the background expansion only.

Next we define the quantity related to the velocity perturbation that gives rise to redshift space distortion:

\begin{equation}
f=-\frac{k^2 v_{m}}{{\mathcal{H}}\Delta_{m}}.
\end{equation}

\noindent
In figure 3, we show the deviation in $f$ from $\Lambda$CDM for various scalar field potential. As in $\Delta_{m}$, here also the deviation is not large and is scale independent for most of the redshifts showing that the deviation is due to the background expansion only.

Hence the effects of dark energy perturbation in comoving density contrast $\Delta_{m}$ and in redshift space distortion parameter $f$ are negligible at all scales. Subsequently for $\Delta_{m}$ and $f$, the difference between $\Lambda$CDM model and any scalar field model are mostly dominated by their difference in background evolution.

\section{The observed galaxy power spectrum}

Distribution of galaxies in our universe is one of the best possible probes to study the evolution of our universe and its contents. Observed galaxy distribution is related to the underlying dark matter distribution and hence by observing various features in the galaxy distribution at different scales, one can study the growth in dark matter fluctuations which in turn can help us to distinguish various dark energy and modified gravity models.

In late eighties, Kaiser \citep{Kaiser:1987qv} argued that we do not see galaxies in real space but in redshift space and its distribution in redshift space is influenced also by the peculiar velocities of the galaxies in addition to the dark matter fluctuations. This gives rise to what is known as the Kaiser redshift space distortion, a measure of the large scale velocity field. This contains valuable information about the underlying cosmology.

In addition, the observed galaxy distribution is also affected by gravitational lensing through an effect known as magnification bias \citep{Moessner:1997qs}, allowing the faint galaxies to be detected through the magnification due to lensing effect. This depends on the gravitational potentials in the metric integrated along the photon geodesics.

In recent years, people have shown the presence of other effects in the observed galaxy distribution on larger scales. These are purely general relativistic effects and depends on how the gravitational potential, velocity fields as well as the matter density affect the observed number density of galaxies on large cosmological scales \citep{Yoo:2009au,Bonvin:2011bg,Bonvin:2014owa,Challinor:2011bk,Jeong:2011as,Yoo:2012se,Bertacca:2012tp,Duniya:2016ibg,Duniya:2015dpa}. On sub horizon scales, these GR effects are negligible in comparison to other effects like redshift space distortion. However on large cosmological scales, where one needs to consider these GR effects as they can be important in distinguishing different dark energy models.

In a galaxy survey, we observe the fluctuations in the number of galaxies across the sky and at different redshifts and angles. The galaxy number overdensity  $\Delta^{obs}$ is given by \citep{Duniya:2016ibg,Duniya:2013eta,Duniya:2015nva,Challinor:2011bk}

\begin{equation}
\Delta^{obs} = \left[{b + f \mu^2} + \mathcal{A} (\frac{\mathcal{H}}{k})^2 +  i\mu\mathcal{B} (\frac{\mathcal{H}}{k})\right]\Delta_{m},
\end{equation}

\noindent
where $b$ is the scale independent bias on linear scales that relates the dark matter density contrast to the galaxy density contrast, $f$ the redshift space distortion parameter that is defined in the previous section, $\mu  = -\frac{\vec{n}.\vec{k}}{k}$ with $\vec{n}$ denotes the direction of observation$, \vec{k}$ being the wave vector and $k$ being the wavenumber. The parameters $\mathcal{A}$ and $\mathcal{B}$  which are related to the GR corrections, are defined as:
 
\begin{equation}
\mathcal{A} = 3f + (\frac{k}{\mathcal{H}})^2 \Big{[} 3 + \frac{\mathcal{H}'}{\mathcal{H}^{2}} + \frac{\Phi'}{\mathcal{H} \Phi} \Big{]} \frac{\Phi}{\Delta_{m}},
\end{equation}

\begin{equation}
\mathcal{B} = - \Big{[} 2 + \frac{\mathcal{H}'}{\mathcal{H}^{2}} \Big{]} f.
\end{equation}

\noindent
Here we assume the magnification bias to be equal unity \citep{Duniya:2015nva} and also assume a constant comoving galaxy number density for which the galaxy evolution bias is zero. We assume the scale independent linear bias $b=1$ throughout our calculations. We also neglect the time-delay, ISW and weak lensing integrated terms. The first term inside square bracket in equation (21) is related to galaxy bias, the second term is the Kaiser redshift space distortion term.  ${\mathcal{A}}$ is related to the peculiar velocity potential and the gravitation potential whereas ${\mathcal{B}}$ is related to the Doppler effect. Using (21), one can now write the power spectrum for the observed galaxy overdensity (the real part) as \citep{Duniya:2016ibg,Duniya:2013eta,Duniya:2015nva,Jeong:2011as}:

\begin{equation}
P(k,z)= \left[(b + f \mu^{2})^{2} + 2 (b + f \mu^{2}) \Big{(} \frac{\mathcal{A}}{y^{2}} \Big{)} + \frac{\mathcal{A}^{2}}{y^{4}} + \mu^{2} \Big{(} \frac{\mathcal{B}^{2}}{y^{2}} \Big{)}\right]P_{s}(k,z),
\end{equation}

\noindent
where $y = \frac{k}{\mathcal{H}}$ and $P_{s}$ is the standard matter power spectra given by

\begin{equation}
P_{s}(k,z) = A k^{n_{s}-4} T(k)^{2}\left(\frac{|\Delta_{m}(k,z)|}{|\Phi(k,0)|}\right)^{2}.
\end{equation}

\noindent
In the above equation for $P_{s}$, we fix the constant $A$ using $\sigma_{8}$ normalisation and we use the Eisenstein-Hu transfer function for $T(k)$ \citep{Eisenstein:1997ik}. In figure 5, we plot the line of sight ($\mu =1$) observed galaxy power spectrum given by equation (24) for the linear potential only. We assume the spectral index for the initial power spectrum $n_{s} = 0.96$, $\sigma_{8} = 0.8$, $h=0.7$ and $\Omega_{b0} = 0.05$ for our calculation. The behaviour of the total power spectrum clearly shows the enhancement of power on larger scales due to GR corrections as compared to the standard matter power spectrum or when one includes only the Kaiser redshift space distortion term. 

\begin{figure*}
    \centering
    {
        \includegraphics[scale=0.55]{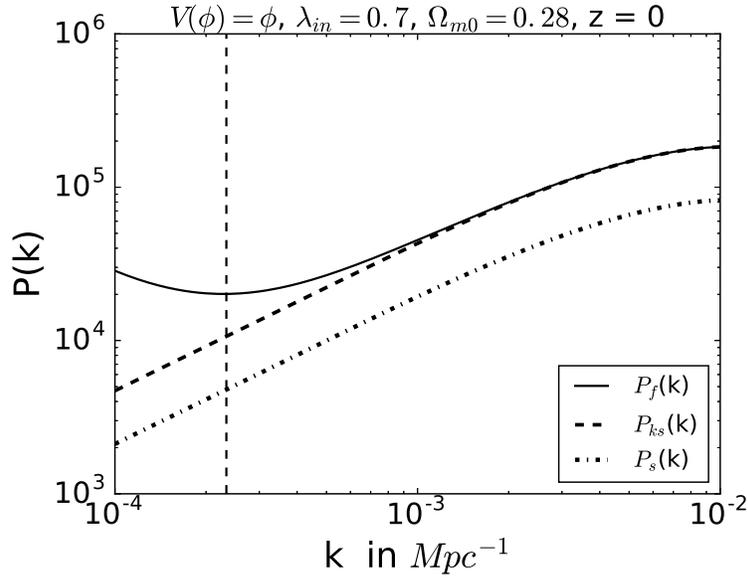}
    }
\caption{Dashed-dotted, dashed and continuous lines are for the usual matter power spectrum $P_{s}$ (given by eqn. (25)), the galaxy power spectrum taking only Kaiser term $P_{ks}$ ( Taking only the first term inside square bracket in eqn. (24)) and the full observed galaxy power spectrum $P$ given by eqn. (24) respectively. The vertical line is the horizon scales at $z=0$.}
\end{figure*}

\begin{center}
\begin{figure*}
\begin{tabular}{c@{\quad}c}
\epsfig{file=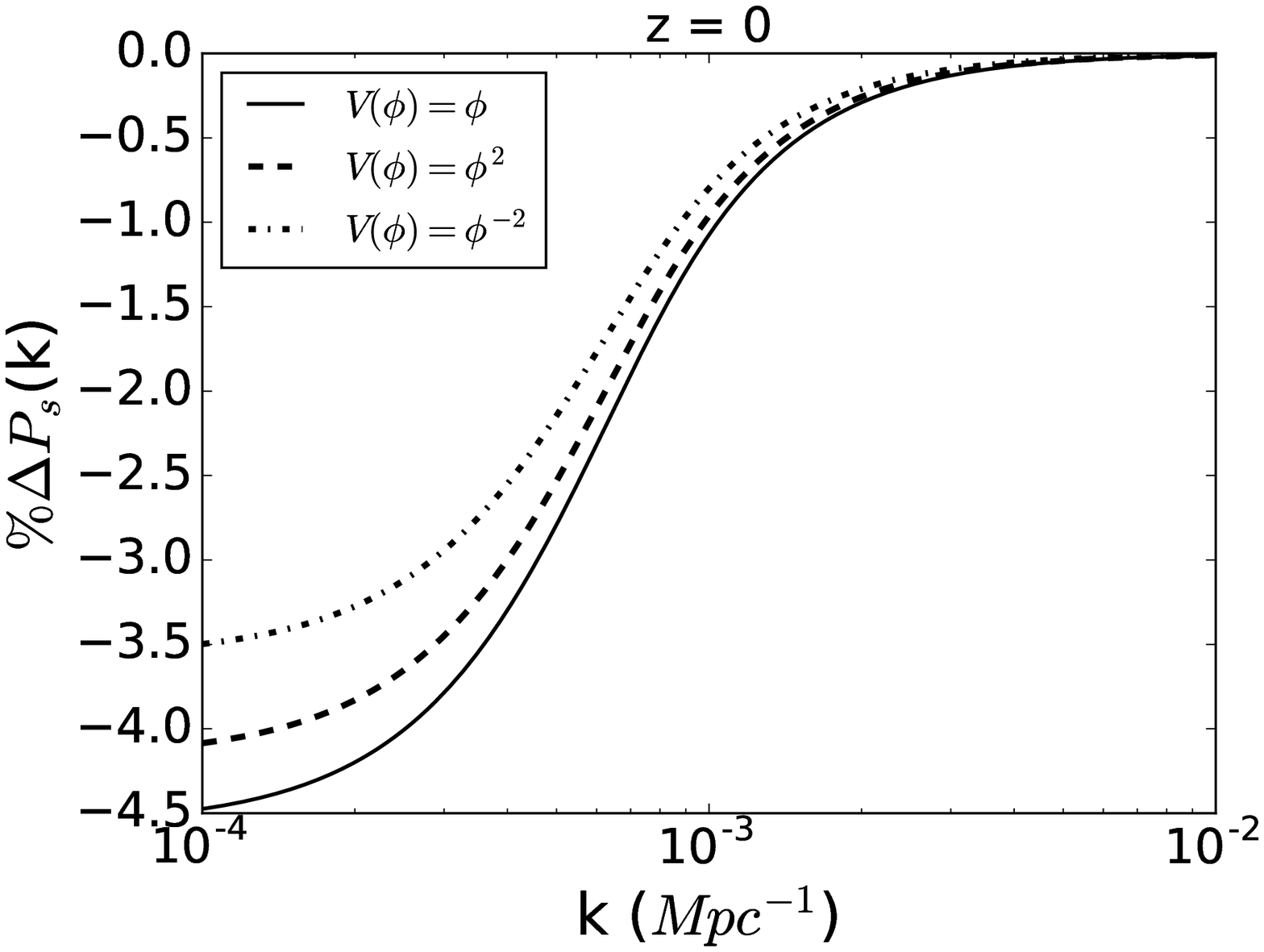,width=6 cm}
\epsfig{file=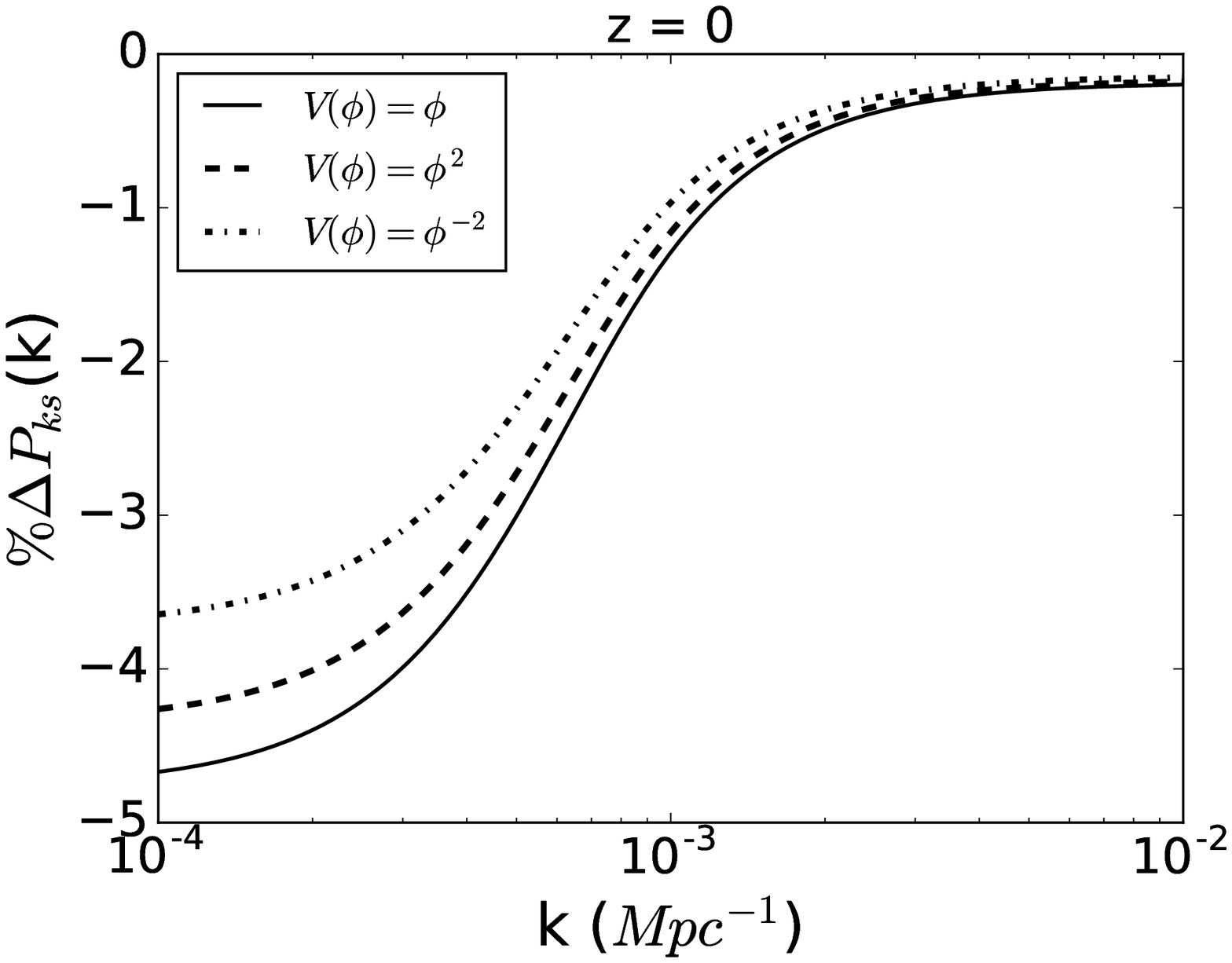,width=6 cm}
\epsfig{file=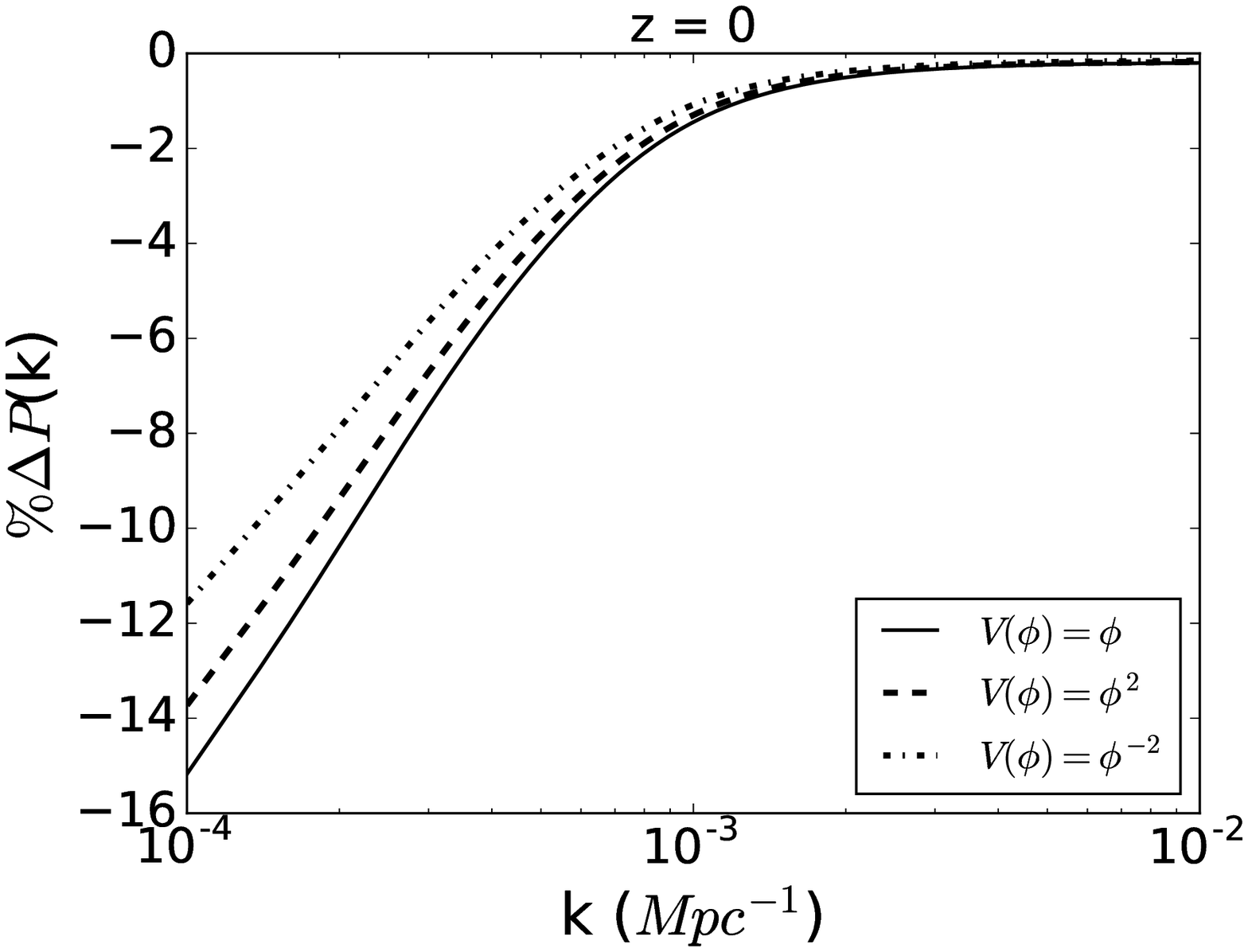,width=6 cm}\\
\epsfig{file=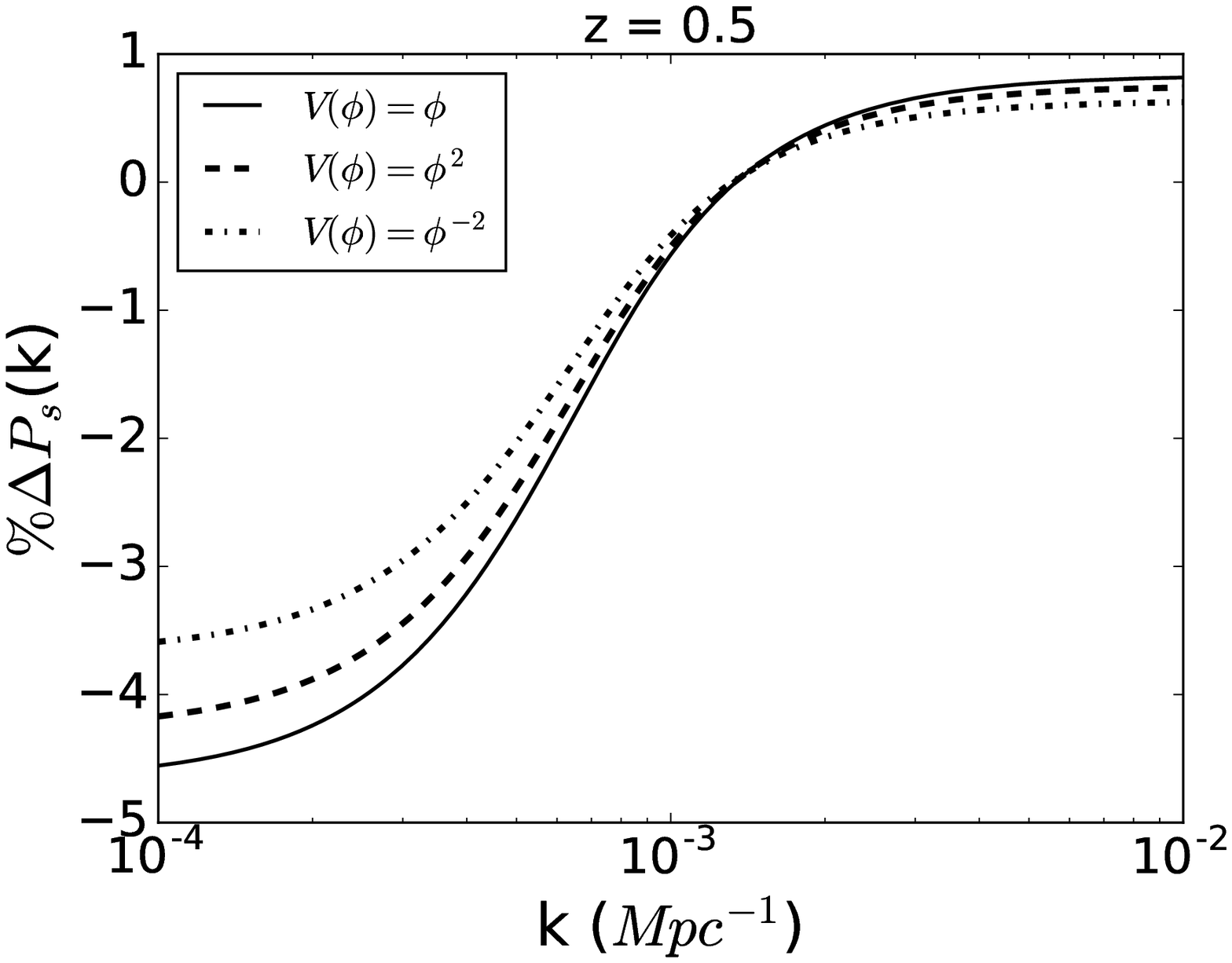,width=6 cm}
\epsfig{file=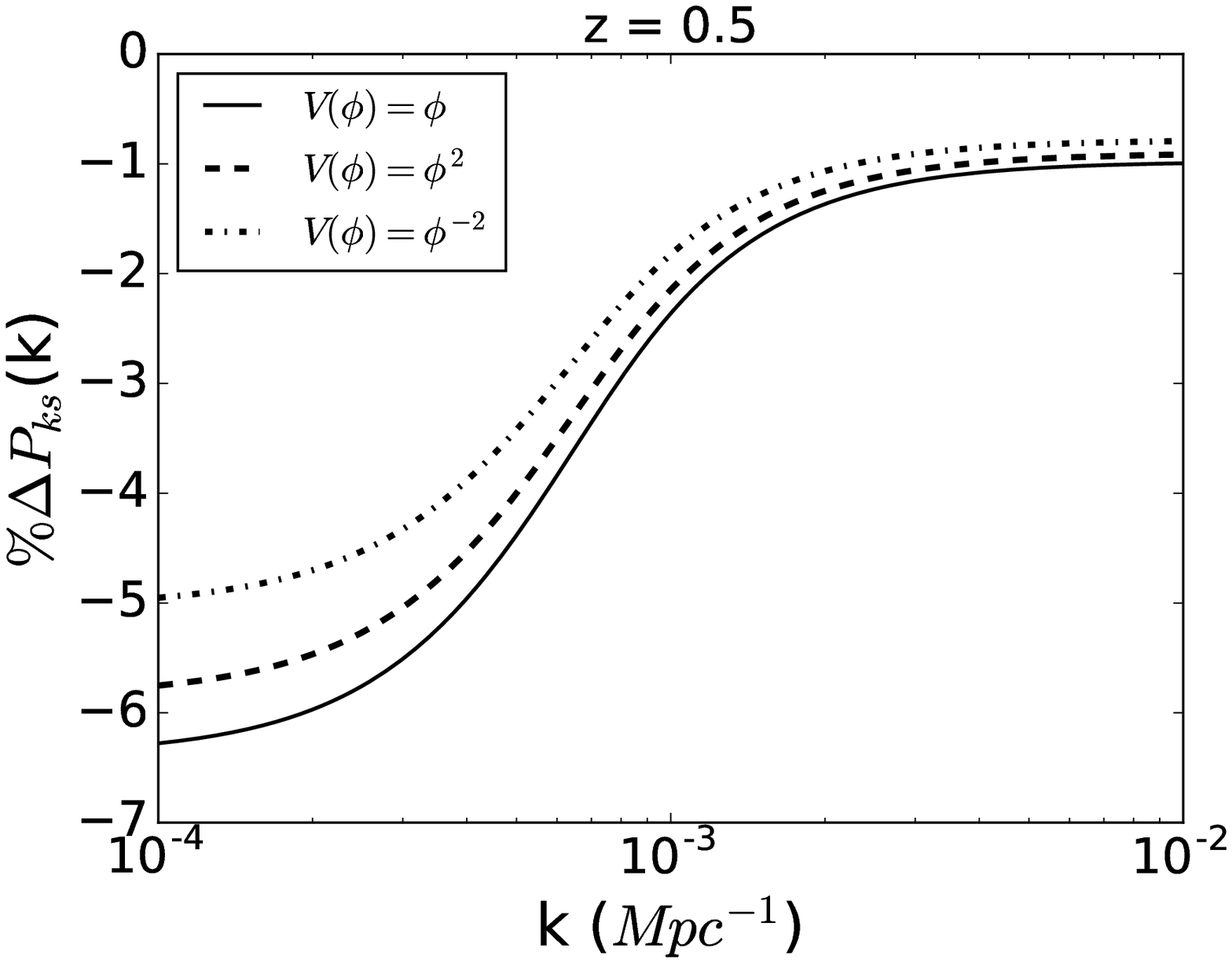,width=6 cm}
\epsfig{file=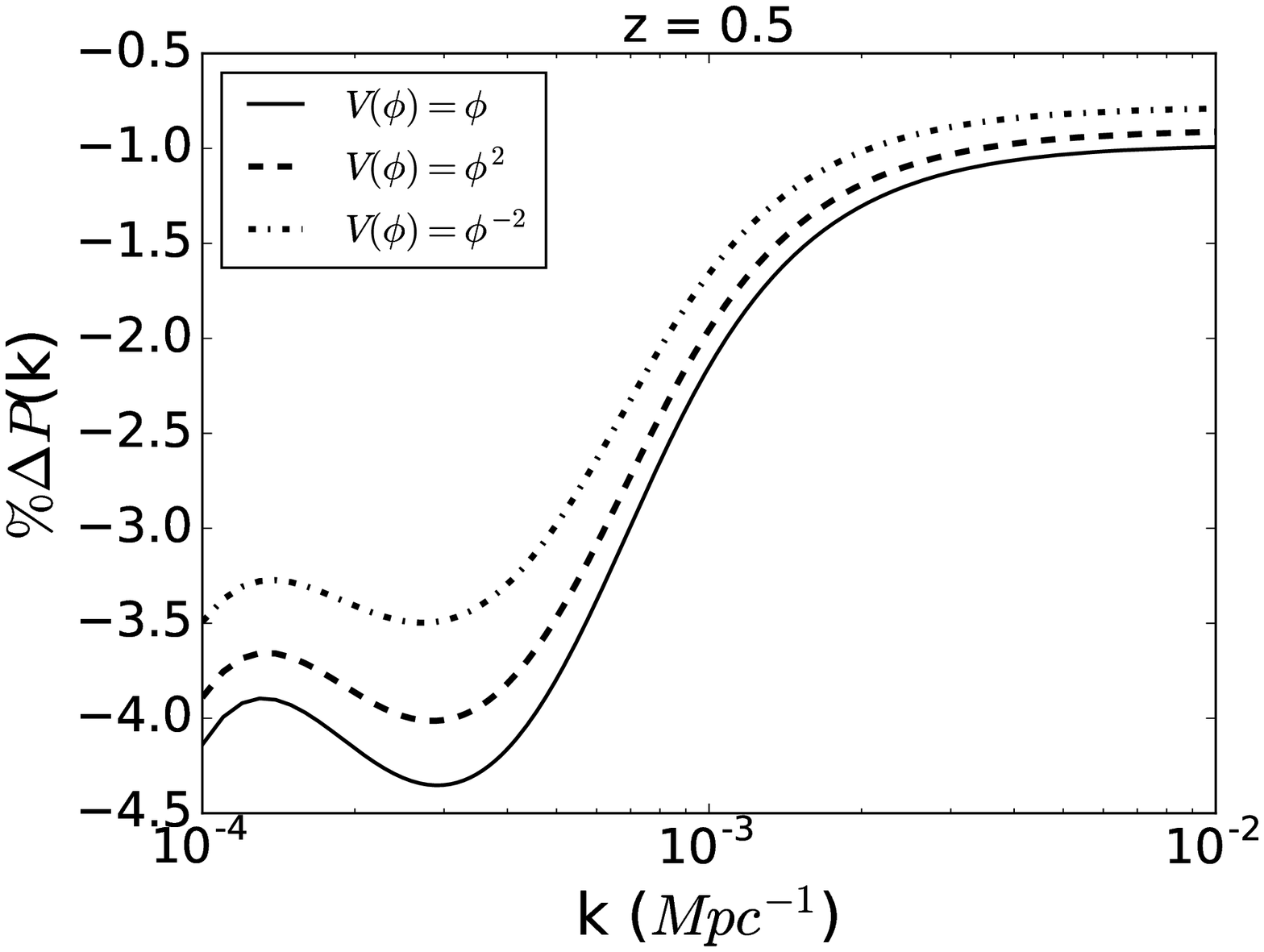,width=6 cm}\\
\epsfig{file=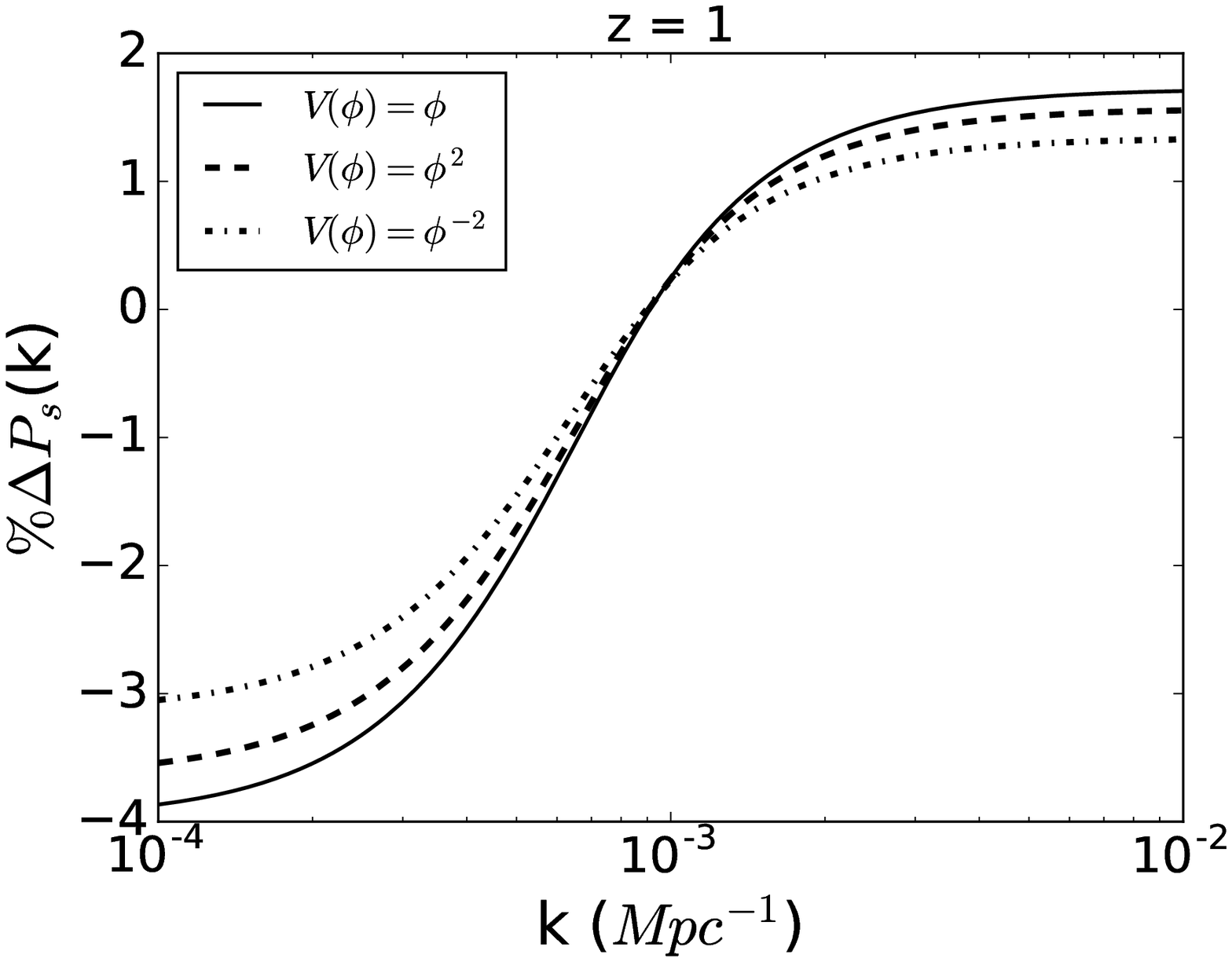,width=6 cm}
\epsfig{file=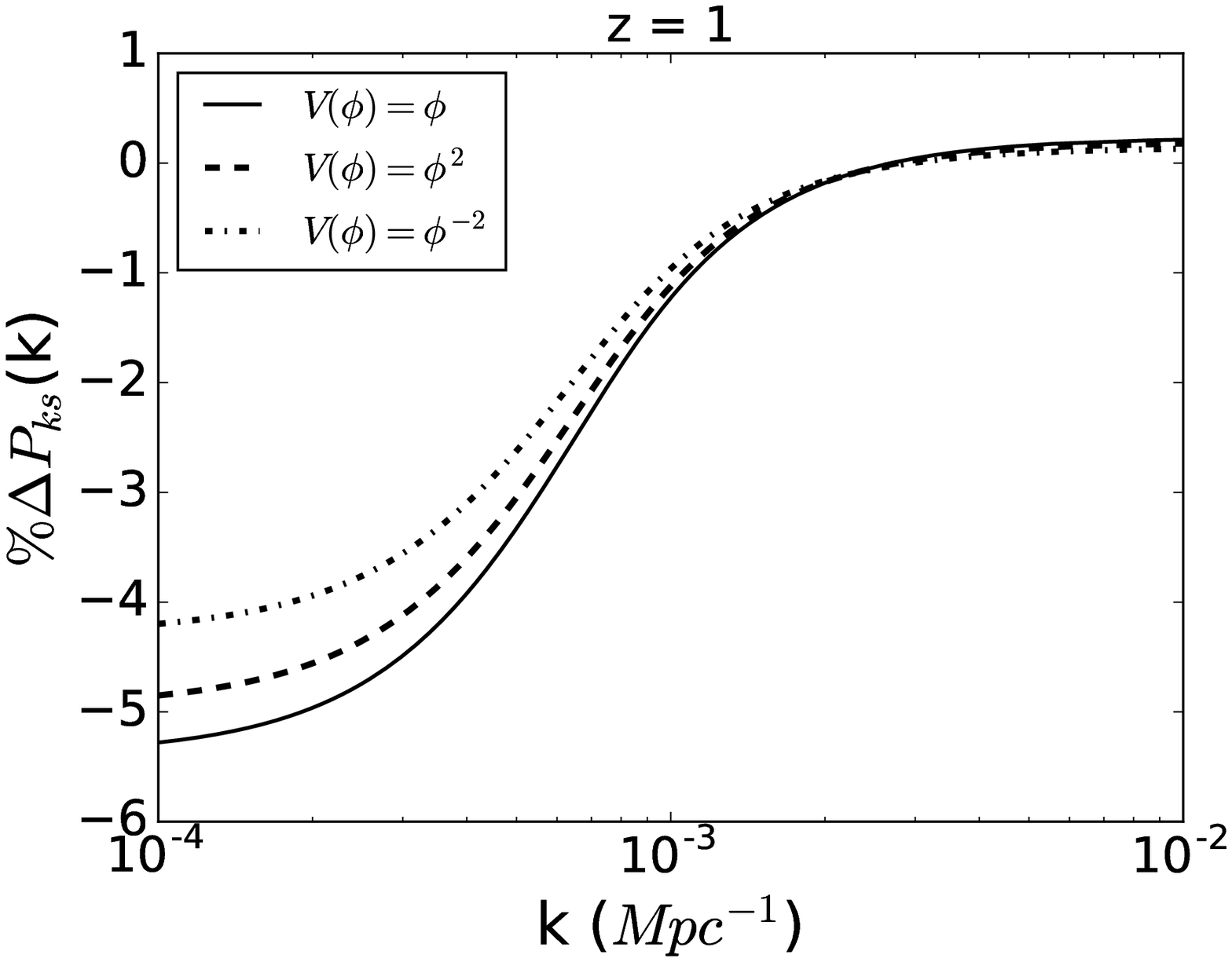,width=6 cm}
\epsfig{file=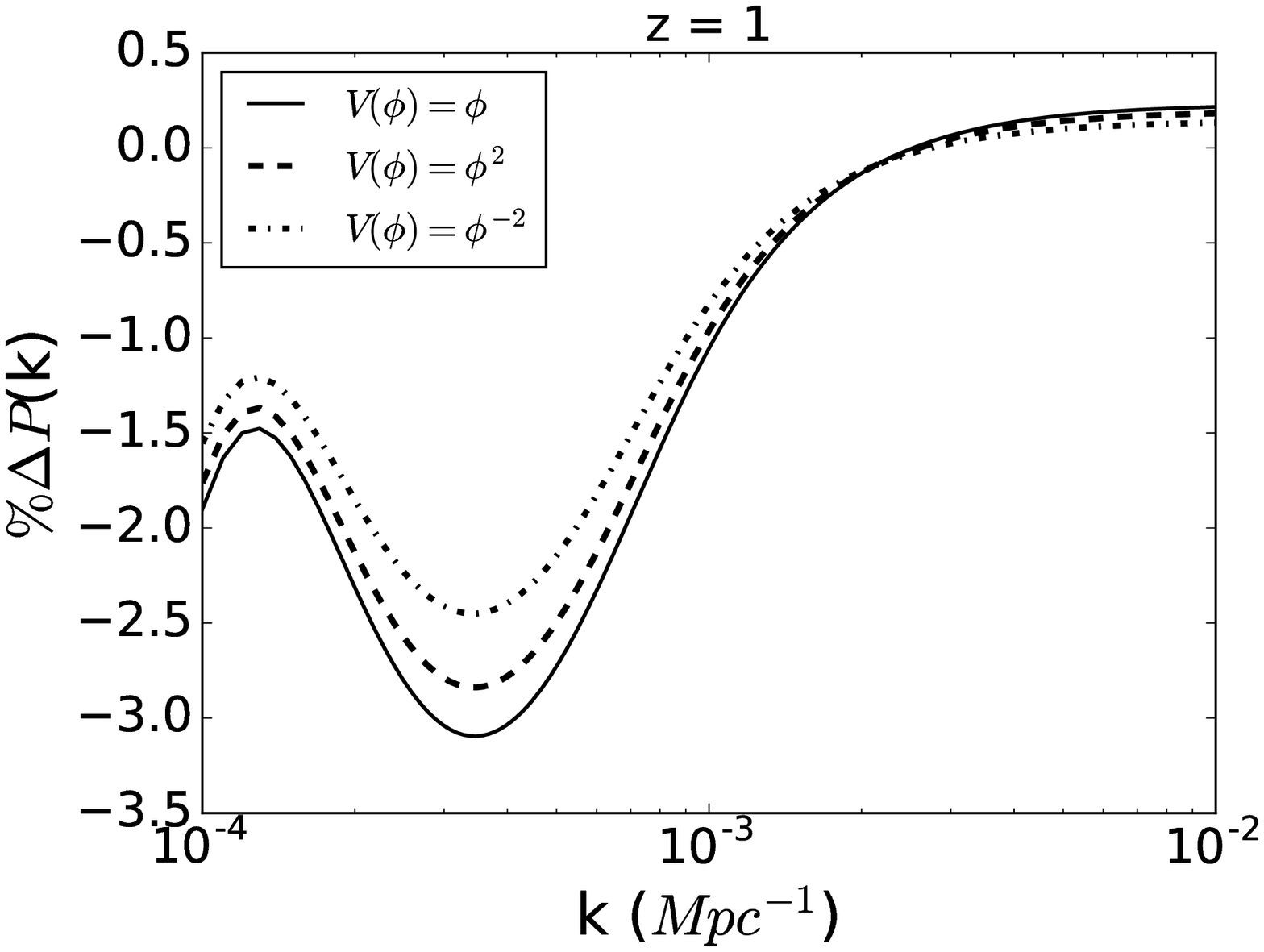,width=6 cm}\\
\epsfig{file=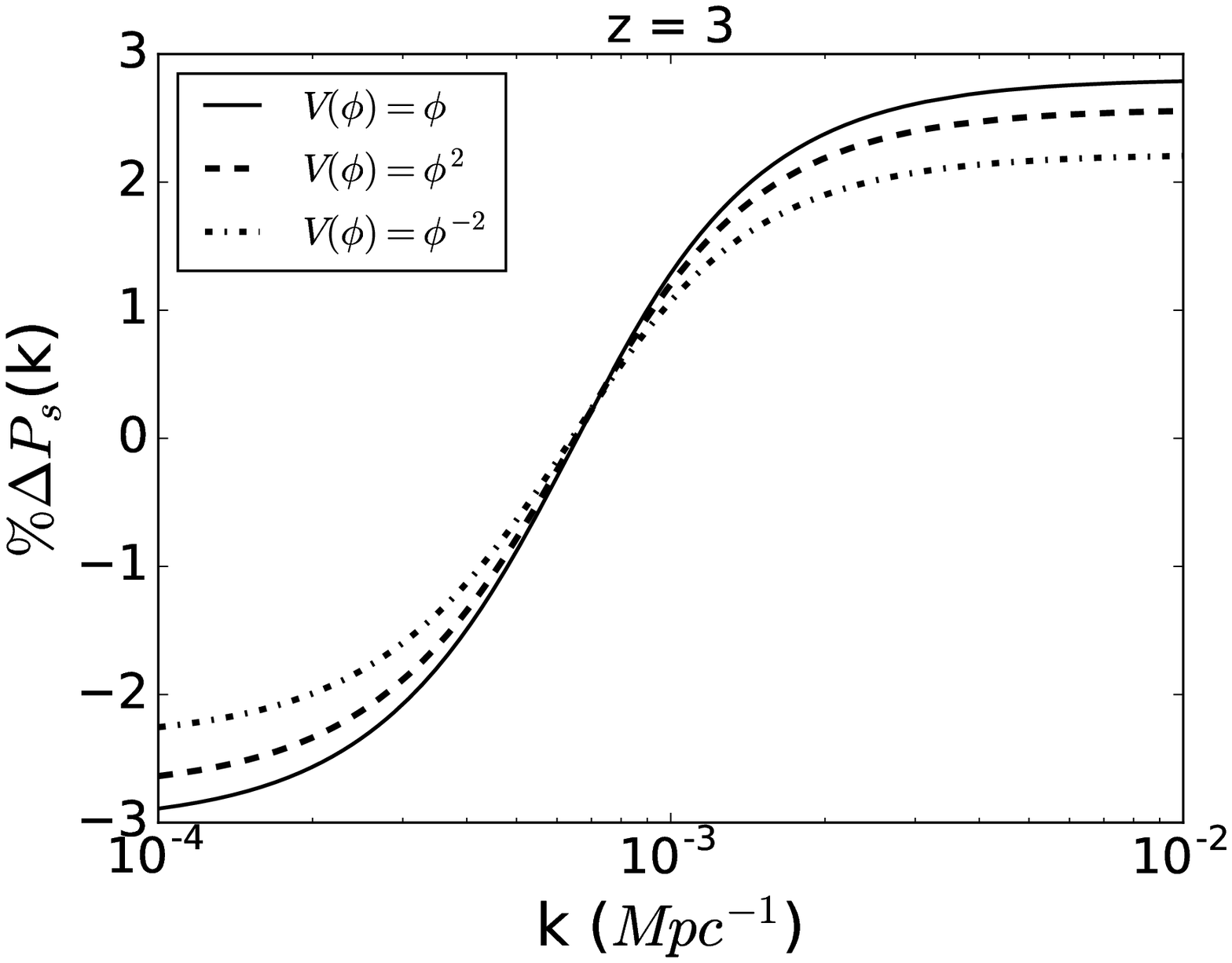,width=6 cm}
\epsfig{file=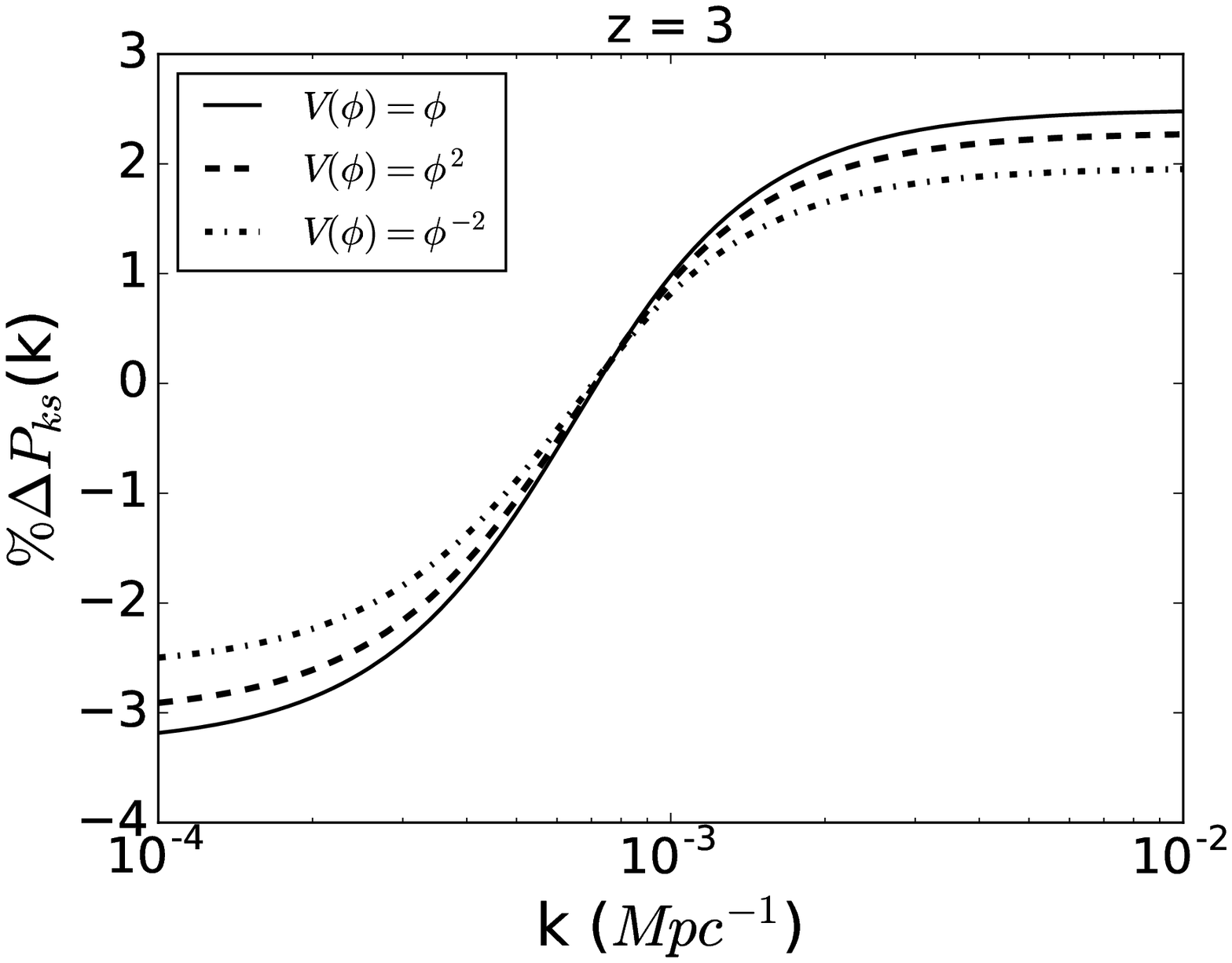,width=6 cm}
\epsfig{file=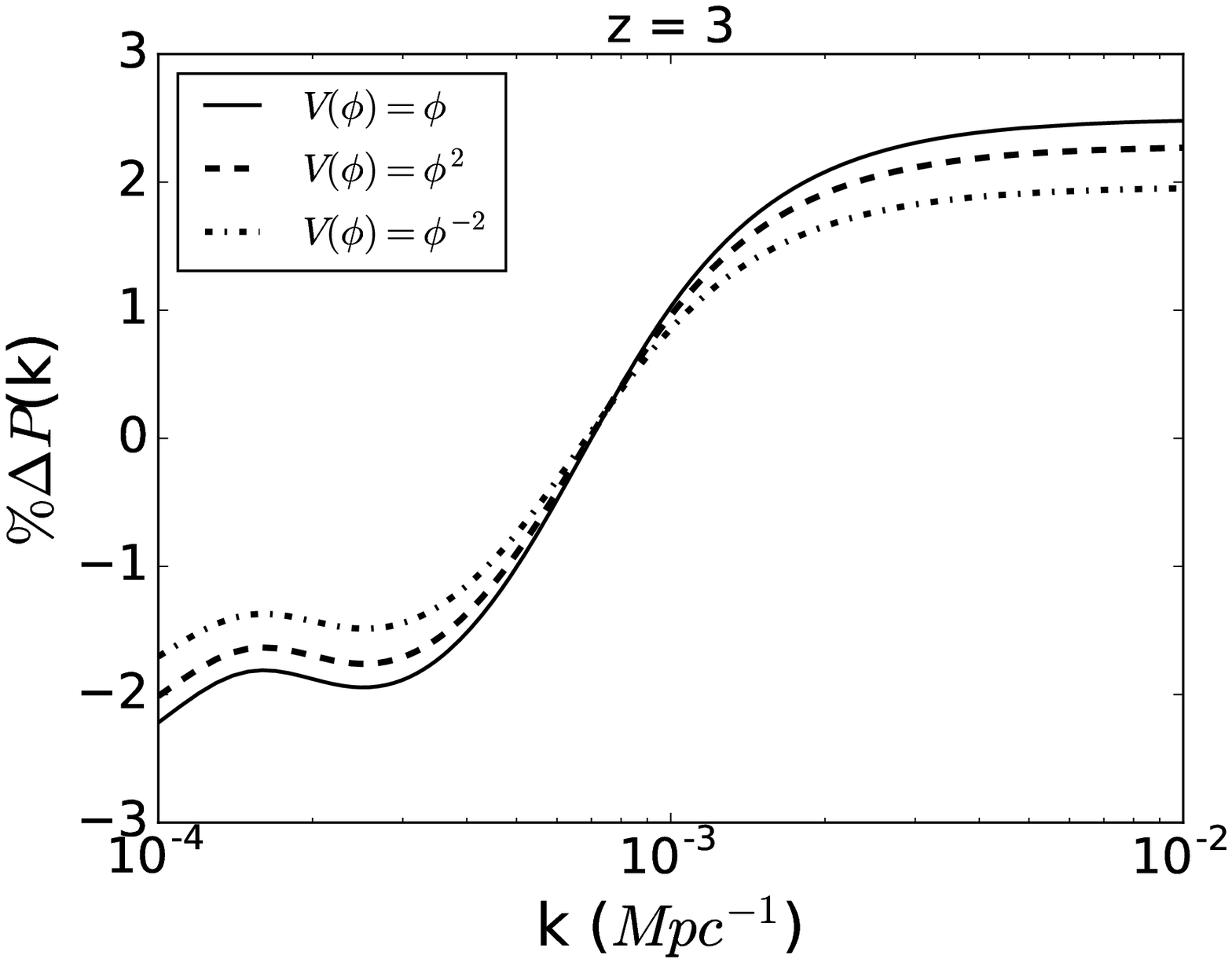,width=6 cm}\\
\end{tabular}
\caption{Percentage deviation in $ P (k) $ from $ \Lambda$CDM model for different potentials as a function of $K$ for four  different redshifts: negative values
in y-axis means they are all suppressed from $ \Lambda $CDM. Left most column is standard matter power spectra $P_{s}$ given by eqn. (25), middle column is for power spectra with Kaiser redshift space distortion term included and the right column is for full observed galaxy power spectra $P$ given by eqn (24) with GR corrections.
}
\end{figure*}
\end{center}

Looking at the expression for observed galaxy overdensity $P$  in eqn. (24) and from our discussion in the previous section, one can get an idea about the effect of dark energy at different scales and redshifts in $P$. The first term under square bracket in (24) is due to Kaiser redshift-space distortion and depends on $f$ (related to the peculiar velocity). We have already shown that the effect of dark energy perturbation in this term is negligible and the only difference that a scalar field dark energy can have from $\Lambda$CDM model through this term is solely due to the difference in background expansion. The fourth term inside the square bracket in (24) is related to Doppler effect and it also depends on $f$ only. Hence in this term also we expect the difference from $\Lambda$CDM solely due to background expansion. The second and third terms inside the square bracket in (24) depend on both $f$ as well as the gravitational potential $\Phi$ and its variation. As $\Phi$ contains extra contribution from dark energy perturbation for large scales and smaller redshift, the effect of dark energy perturbation comes mostly due to these two terms. Moreover in the standard power spectrum $P_{s}$, we have the gravitational potential in the denominator which contains the contribution from the dark energy perturbation on large scales and smaller redshifts. This will also contribute to the deviation from $\Lambda$CDM for scalar field dark energy models.

In figure 6, we show the deviation in different power spectra from the $\Lambda$CDM for different scalar field potentials. We show it for the standard power spectrum $P_{s}$, power spectrum with Kaiser term only $P_{ks}$ and for the full observed galaxy power spectrum $P$ containing GR corrections. We show the deviation as function of scales $k$ for four different redshifts.

From the expression of $P_{s}(k,z)$ in equation (25), one can see that it depends on  $\Delta_{m}$ and $\Phi$. As shown in figure 3, the effect of dark energy perturbation is negligible in $\Delta_{m}$, hence $\Delta_{m}$ is largely dominated by the background expansion. In $\Phi$ there is a contribution from dark energy perturbation on large scales that suppresses $P_{s}$ on large scales in scalar field models compared to $\Lambda$CDM. Keeping this in mind, one can conclude from figure (6) that, for standard power spectra $P_{s}$, on smaller scales the difference from $\Lambda$CDM is dominated by background expansion which increases with redshifts and saturates at $2-3\%$ enhancement from $\Lambda$CDM at redshift $z=3$ and higher. For large scales, there is an extra contribution for dark energy perturbation in the gravitational potential $\Phi$ that suppresses the $P_{s}(k)$ at large scales and this suppression is highest (around $4-5\%$) at smaller redshifts and decreases with redshifts.

When we add the Kaiser redshift distortion term and calculate $P_{ks}(k,z)$, there is an extra effect due to the growth function $f$. But in figure (4), we show that effect of dark energy perturbation is also very small in $f$ and it is mostly dominated by the background expansion only. Due to the added effect of $f$, the overall behaviour shifts slightly up or down depending on the redshifts without any extra $k$-dependence. This is because only background expansion affects $f$ and it does not result any extra k-dependence.

When we add the GR correction terms given in equations (22) and (23) and calculate the full observed galaxy power spectrum $P(k,z)$, there are large suppression at large scales. At smaller redshifts, the suppression from $\Lambda$CDM is now around $12-15\%$ depending on the potentials. Comparing this to difference in $P_{ks}$, one can conclude that the effect of GR corrections alone on large scales is around $9-10\%$. This is indeed a large effect. On smaller scales, the GR corrections are negligible and one can see that the full observed $P(k,z)$ has similar behaviour as in $P_{ks}(k,z)$.

\section{Thawing Vs Tracker}

Till now, we consider the thawing class of scalar field models. As we mention in section (3.1), in thawer models the scalar field is initially frozen due to large Hubble damping resulting the equation of state of the scalar field to be close to $-1$. With expansion of the universe, Hubble damping decreases and the scalar field thaws away from the frozen state and the equation of state of the scalar field slowly increases towards $w > -1$. In future, these models can also exit the accelerating period as the equation of state increases towards more higher values. Hence thawing models can also in principle give transient acceleration.

There is another class of models, known as the {\it tracker models} where initially the scalar field mimics the background matter density ($w_{\phi} \sim 0$) which can be achieved by fast rolling of the field in a steep part of the potential. In late times, the scalar field potential flattens up and the scalar field finally freezes to $w \sim -1$ behaviour. The Large scale structuring in such tracker model has been studied by Duniya et al \citep{Duniya:2013eta}.

In this section, we compare the large scale structuring in thawer and tracker models.  A broad class of potentials can give the thawing behaviour but to have proper tracking behaviour of the scalar field, the form of the potential is very much restricted.  Here we consider a  double exponential type of potential that has been considered earlier \citep{Barreiro:1999zs}:

\begin{equation}
V (\phi) = M^{4} [ e^{- \mu_{1} \frac{\phi}{M_{pl}}} + e^{- \mu_{2} \frac{\phi}{M_{pl}}} ]
\end{equation}

\noindent
where $ M $ is a constant having dimension of mass. The autonomous system of equations in eq. (16) is independent of the value of the $ M $. To get the tracker behaviour for the double exponential potential, we set $ \gamma_{i} = 1 $, $ \lambda_{i} = 19.999985 $ with $ \mu_{1} =20 $ and $ \mu_{2} = 0.1 $;  we assume $ \Omega_{\phi}^{0} = 0.72 $, same as in section 3, to fix $ \Omega_{\phi}^{i}$. The initial conditions for the perturbations are the same as mentioned in section 3.1. For such potential, the equation of state for the scalar field always freezes to the value  $ -1 + \frac{\mu_{2}^{2}}{3} $ in late times. To compare with the thawing behaviour, we take the liner potential for thawer model as discussed in earlier sections as this has the highest deviation from $\Lambda$CDM model as shown in figure 6.

In figure 7 and 8, we compare the behaviour of the equation state and the energy density for the scalar field for tracker and thawer models. As one can see, for the tracker case, the scalar field behaves like matter ($w=0$) in the past and its energy density tracks that of the matter in the past. At late times, it exits the tracking behaviour and starts dominating the matter energy density and its equation of state asymptotically freezes to $w\approx -1$. 

\begin{center}
\begin{figure*}
\begin{tabular}{c@{\quad}c}
\epsfig{file=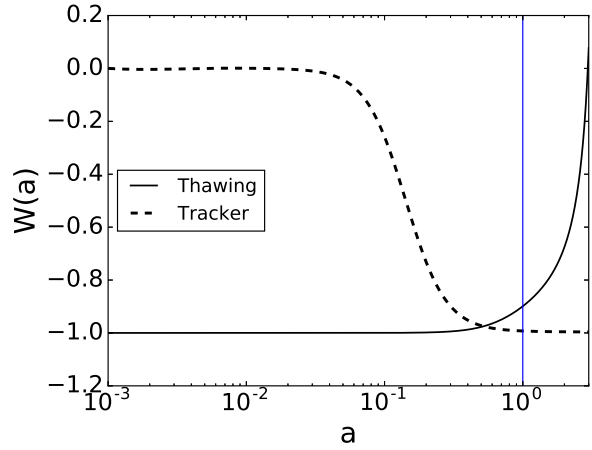,width=9.0 cm}
\end{tabular}
\caption{Behaviour of the Equation of state for the scalar field both for thawing and tracker models respectively. The vertical blue line represents the present time.
}
\end{figure*}
\end{center}

\begin{center}
\begin{figure*}
\begin{tabular}{c@{\quad}c}
\epsfig{file=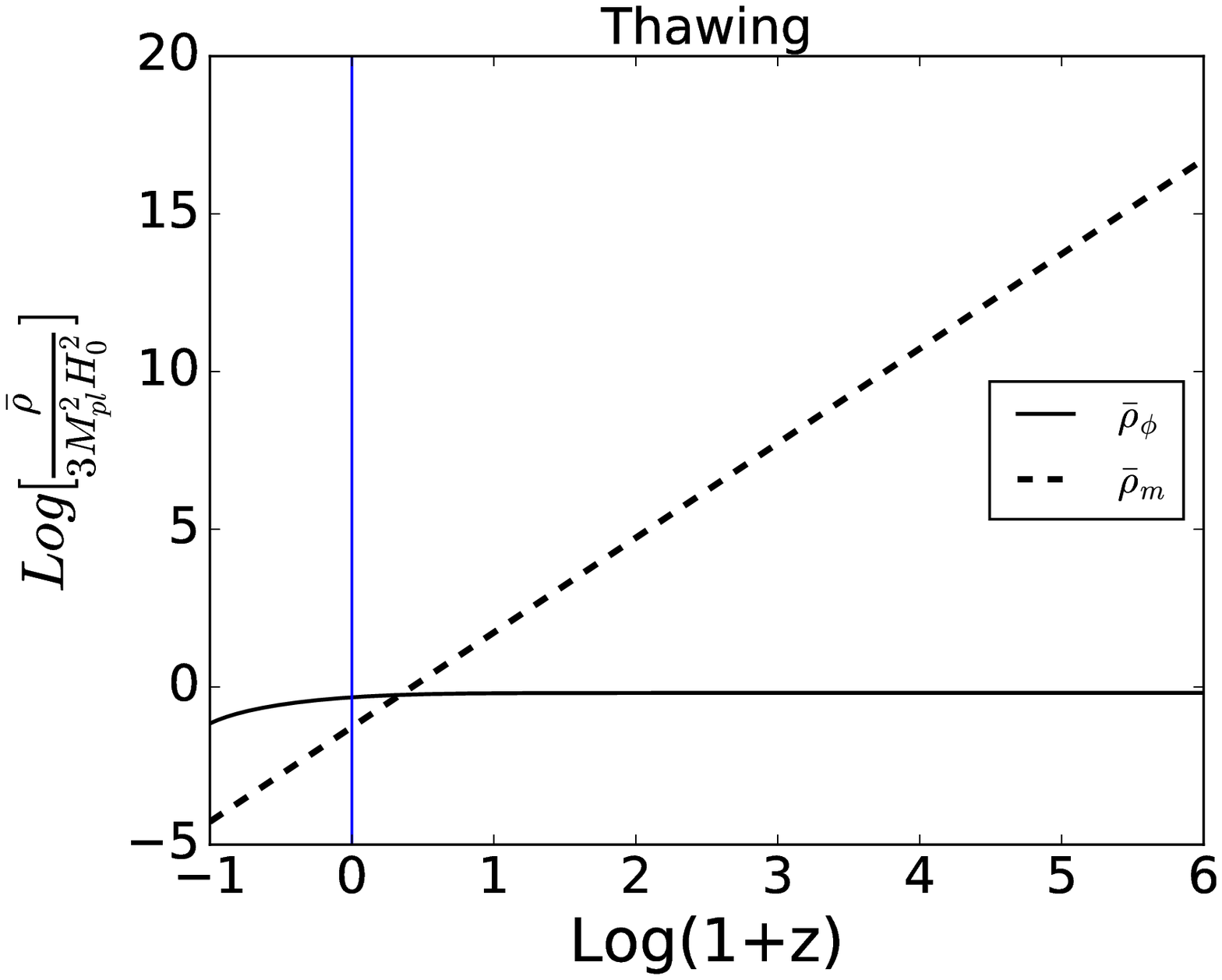,width=8.5 cm}
\epsfig{file=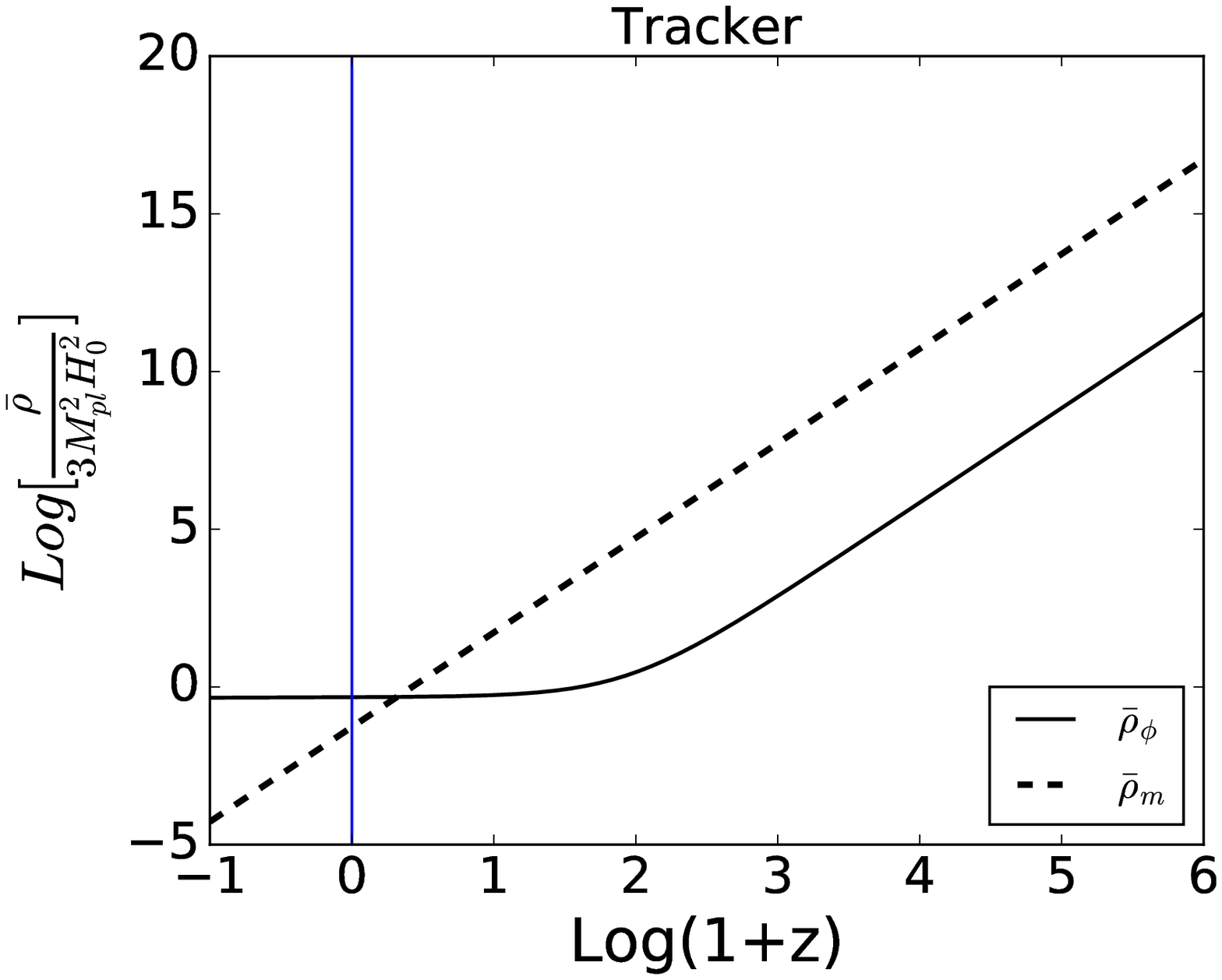,width=8.5 cm}
\end{tabular}
\caption{Behaviour of the energy density of the scalar field and matter energy density. Left panel is for the thawing model and the right panel is for the tracker models. The vertical blue lines represent the present time.
}
\end{figure*}
\end{center}

\begin{center}
\begin{figure*}
\begin{tabular}{c@{\quad}c}
\epsfig{file=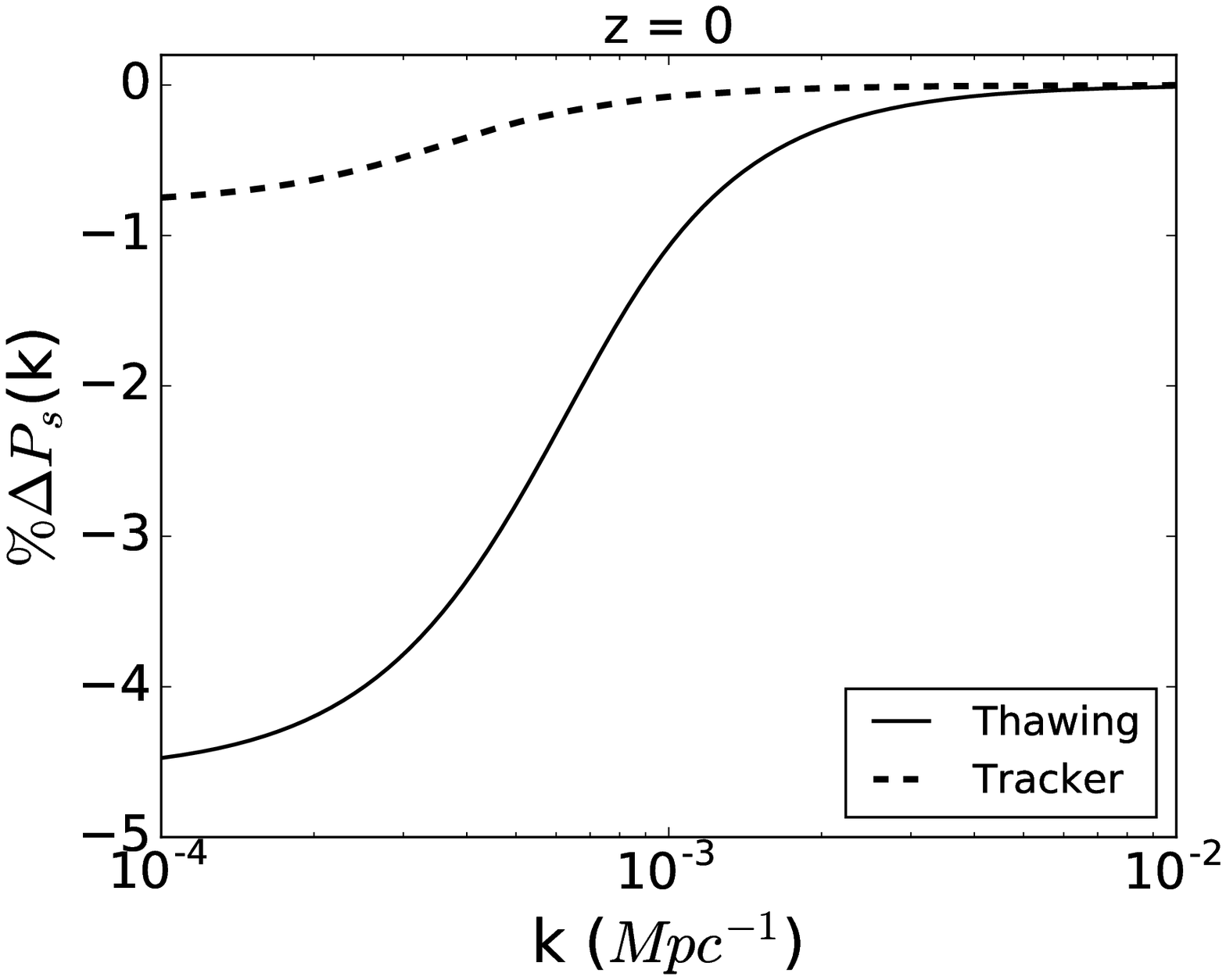,width=5.8 cm}
\epsfig{file=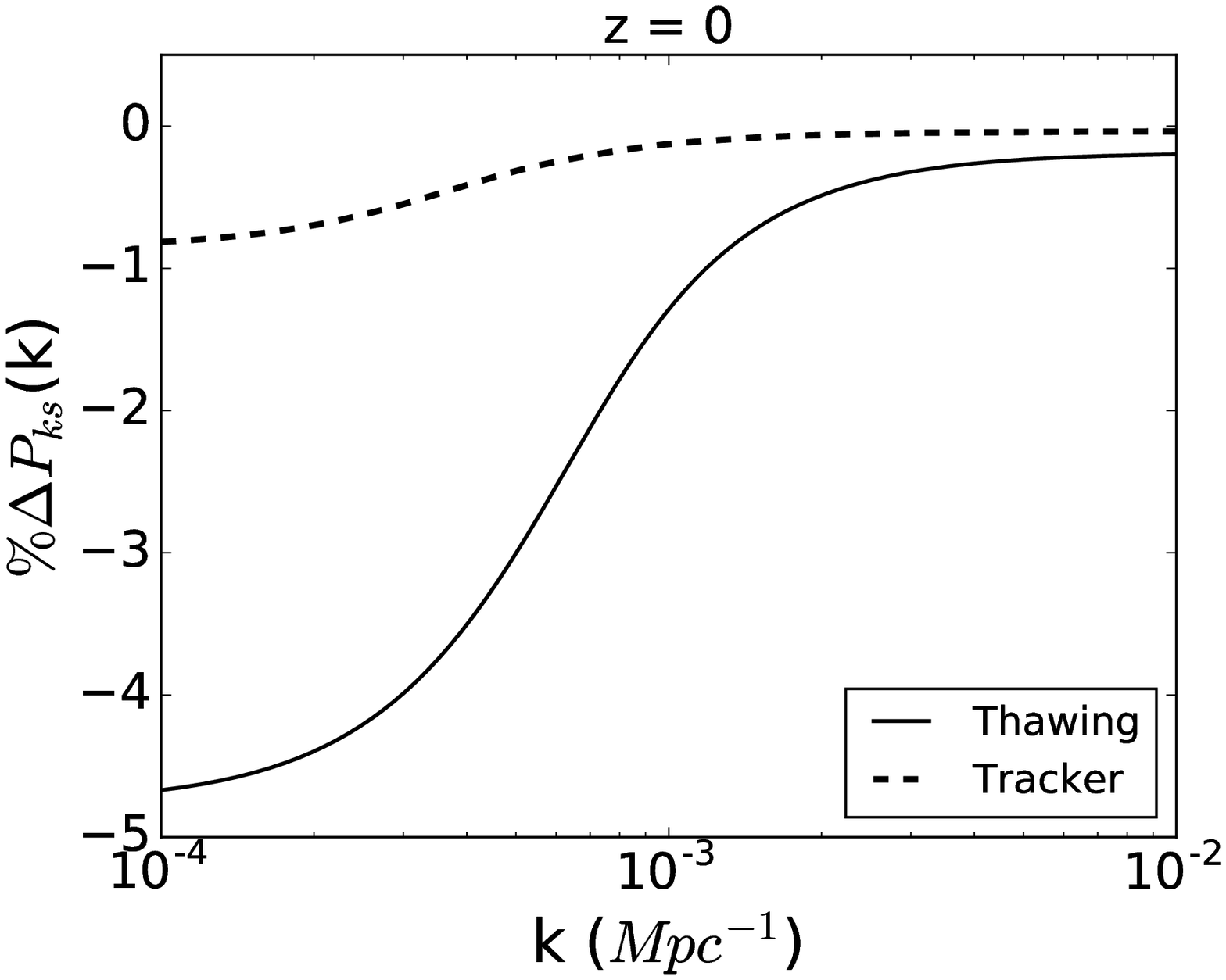,width=5.8 cm}
\epsfig{file=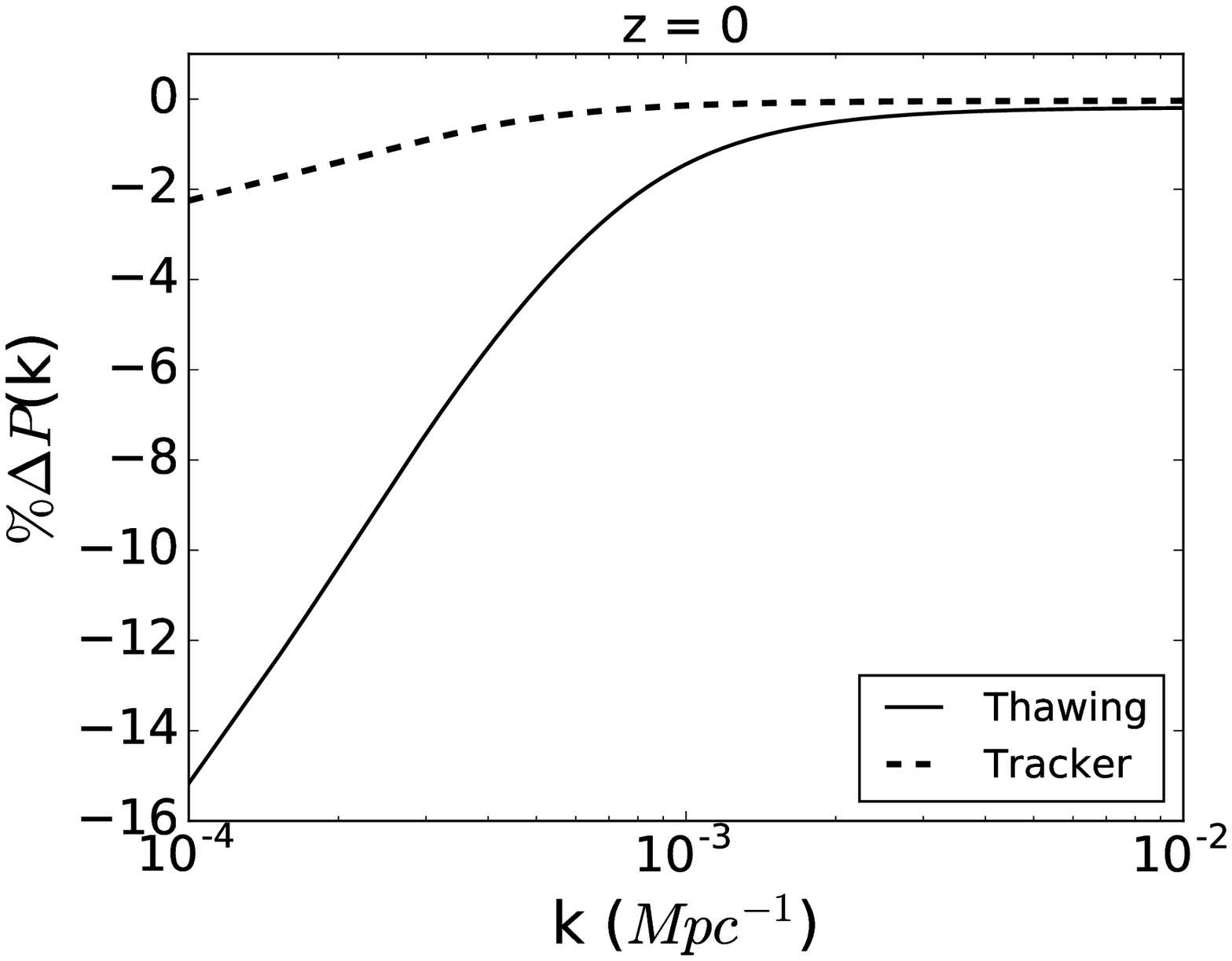,width=5.8 cm}\\
\epsfig{file=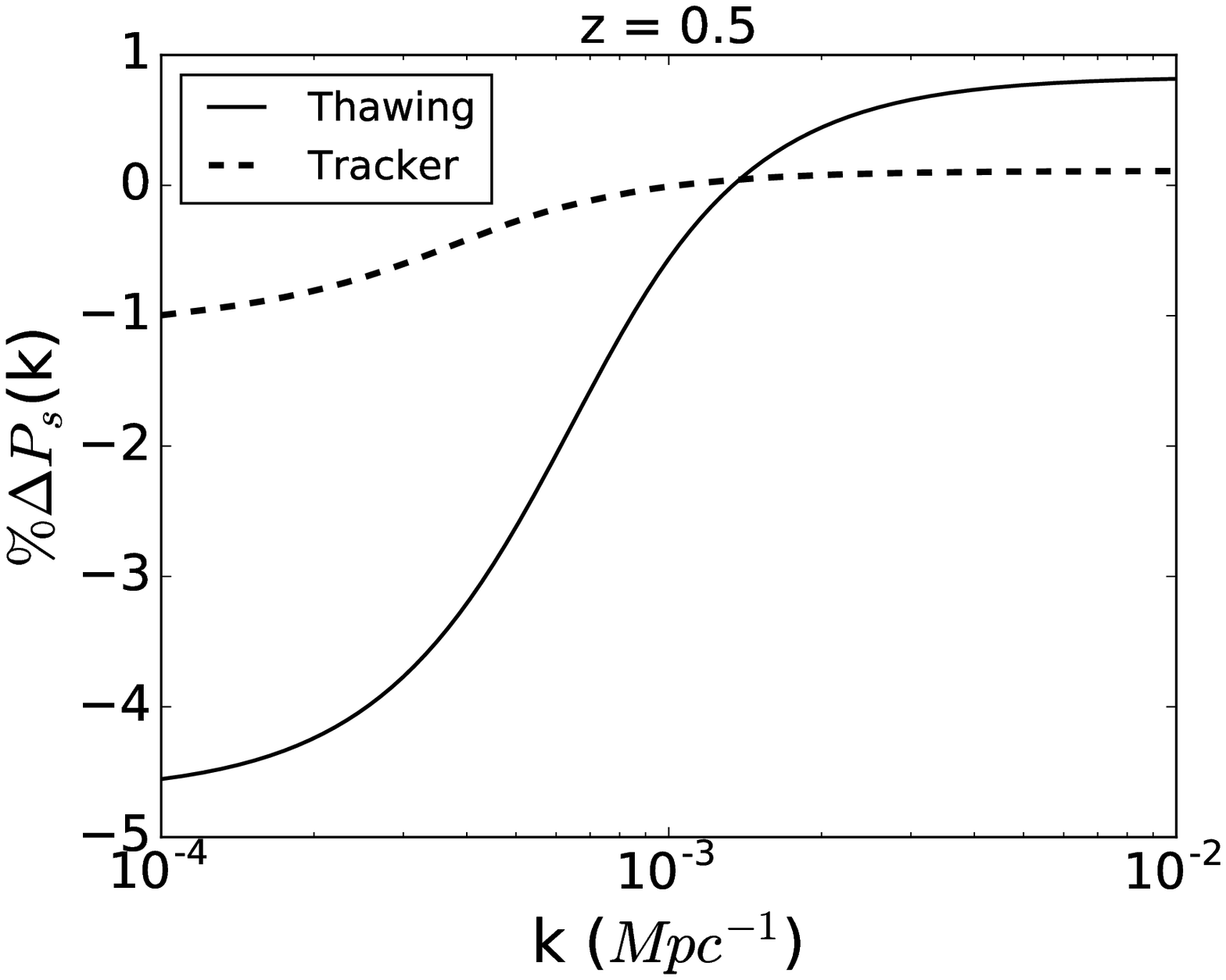,width=5.8 cm}
\epsfig{file=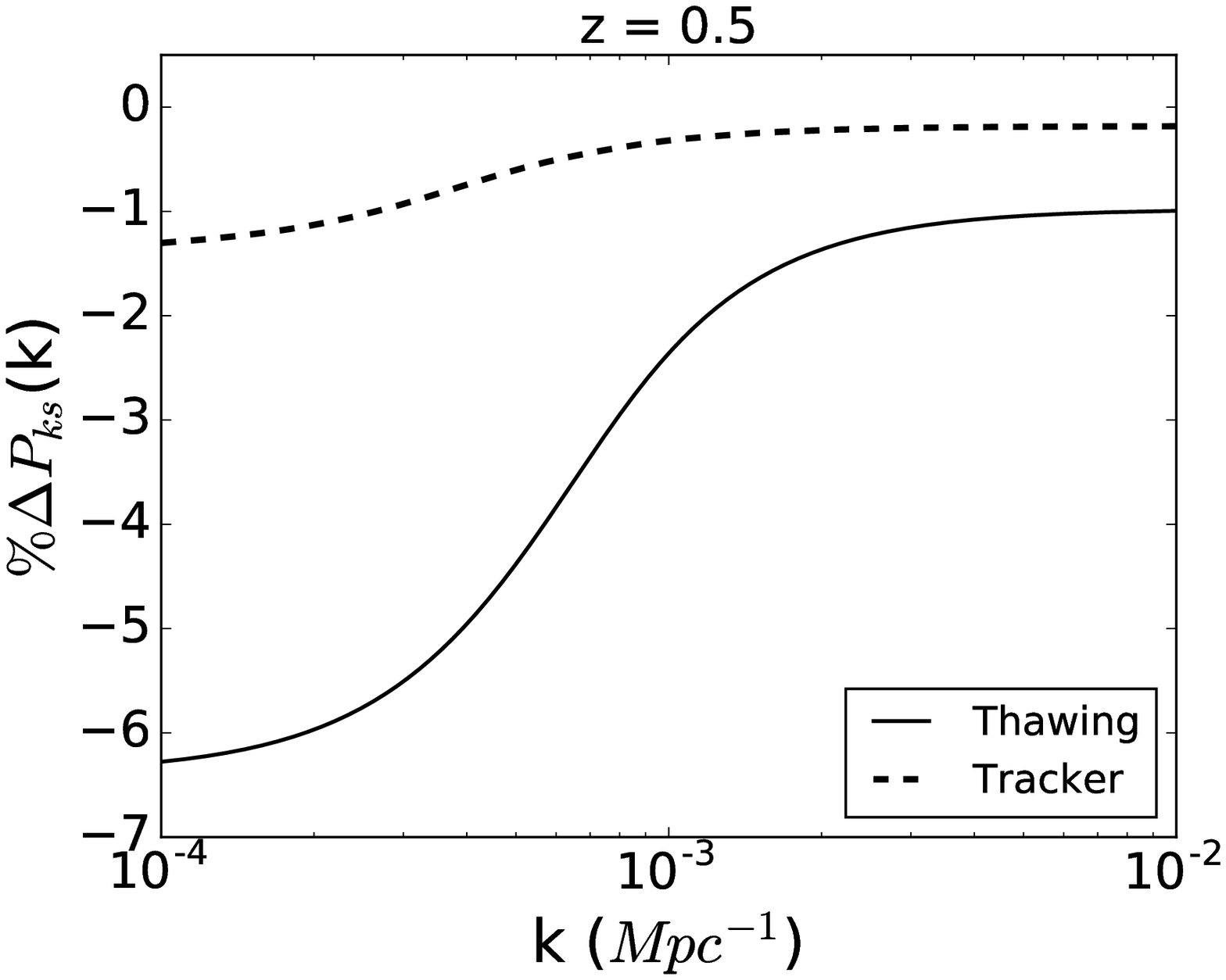,width=5.8 cm}
\epsfig{file=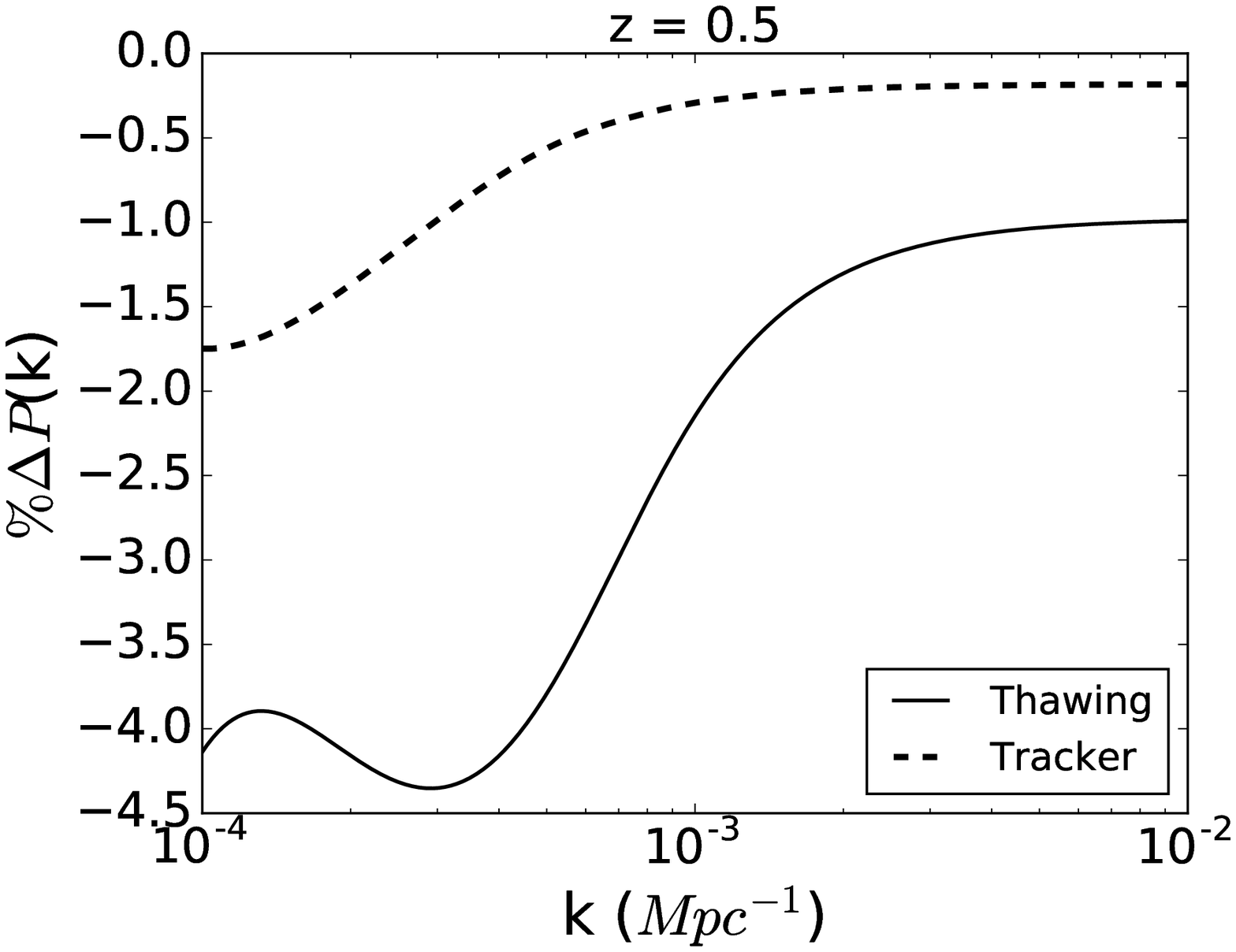,width=5.8 cm}\\
\epsfig{file=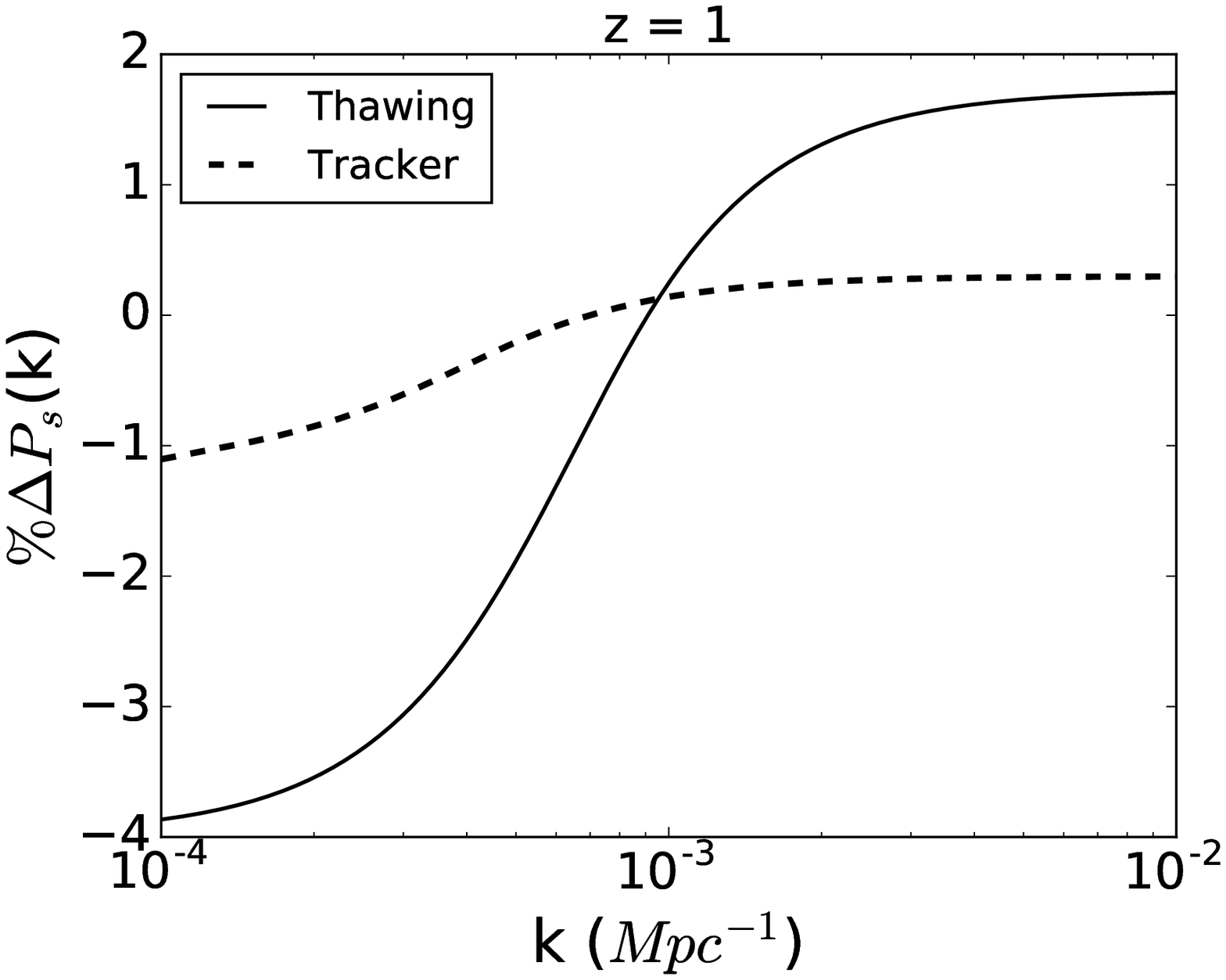,width=5.8 cm}
\epsfig{file=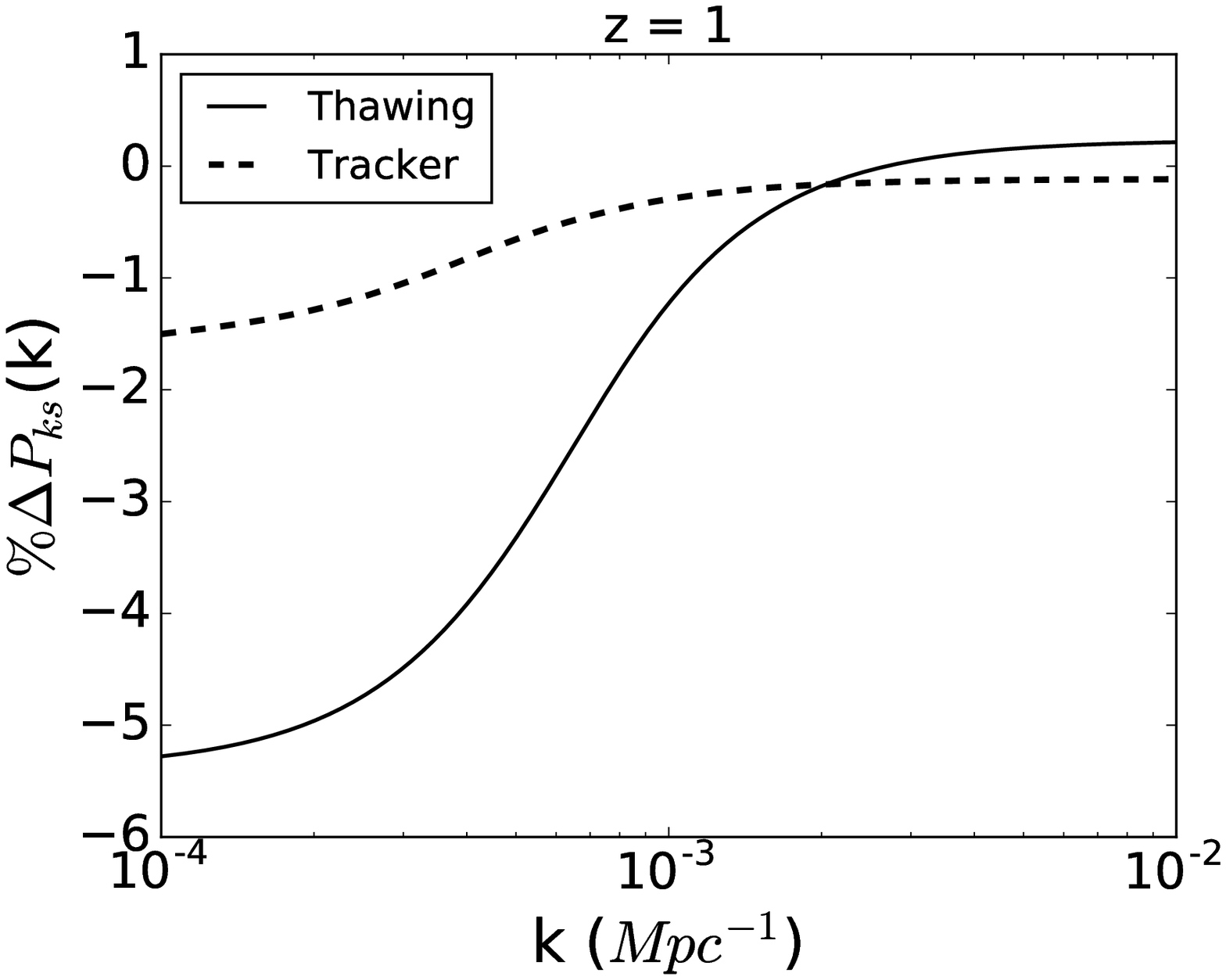,width=5.8 cm}
\epsfig{file=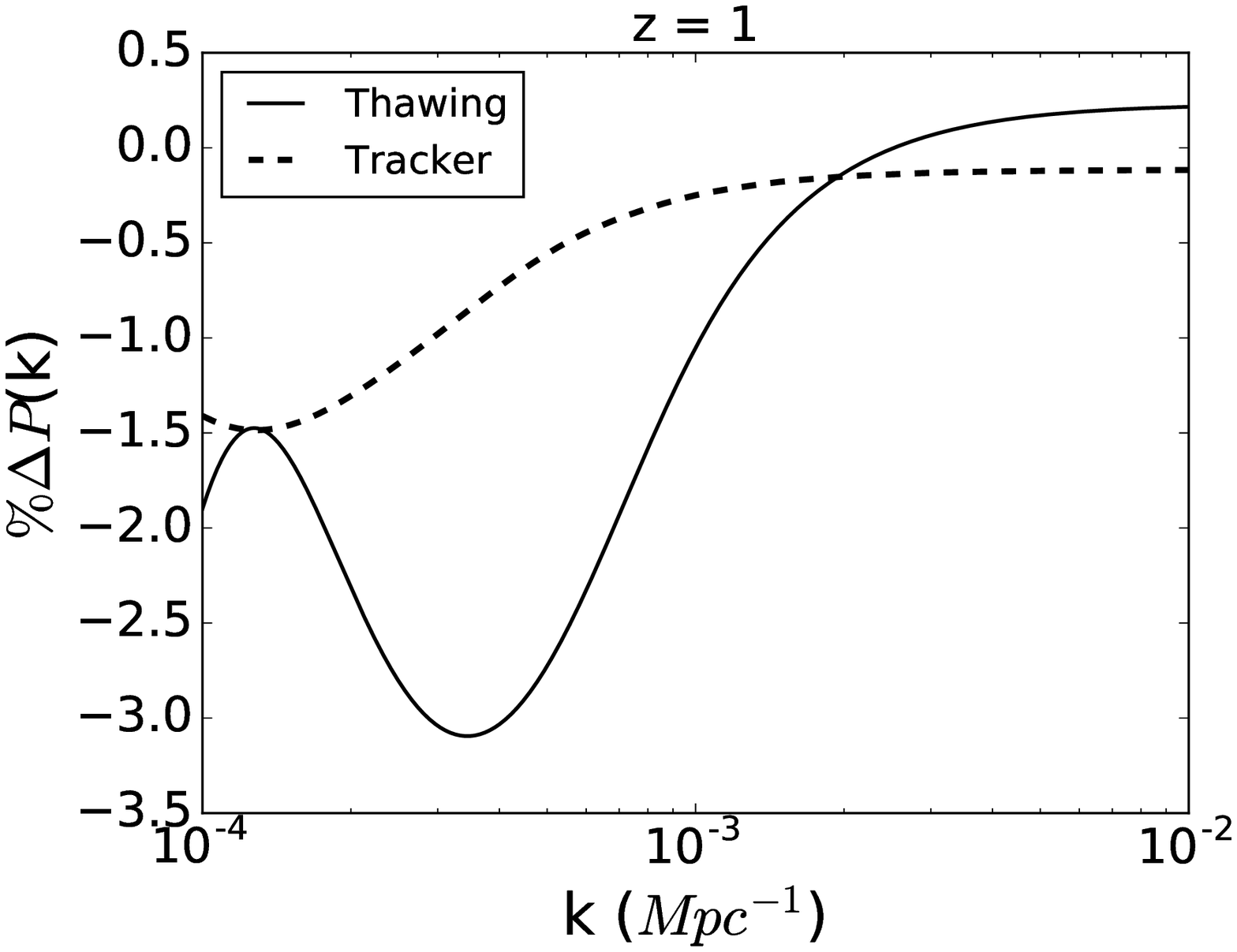,width=5.8 cm}\\
\epsfig{file=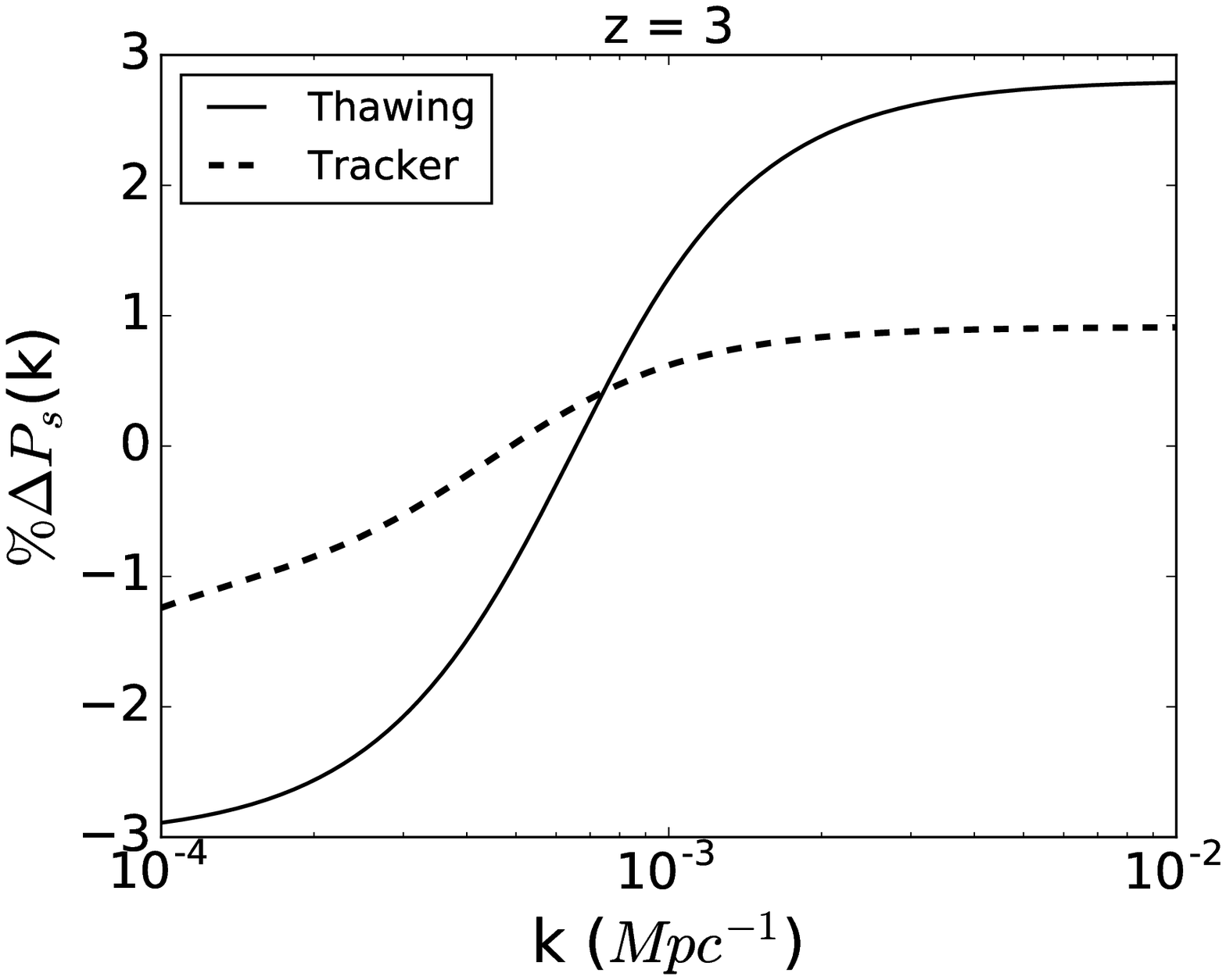,width=5.8 cm}
\epsfig{file=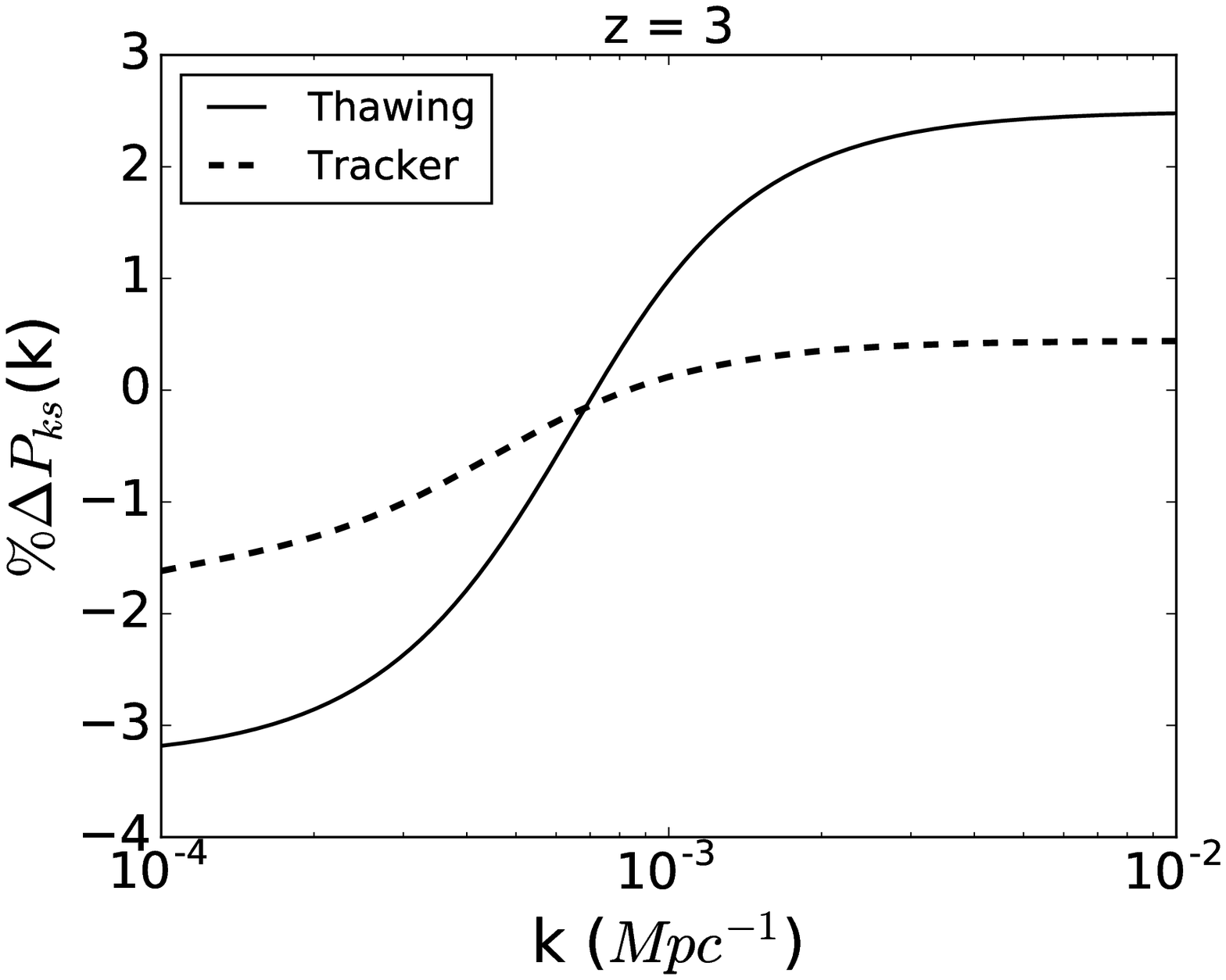,width=5.8 cm}
\epsfig{file=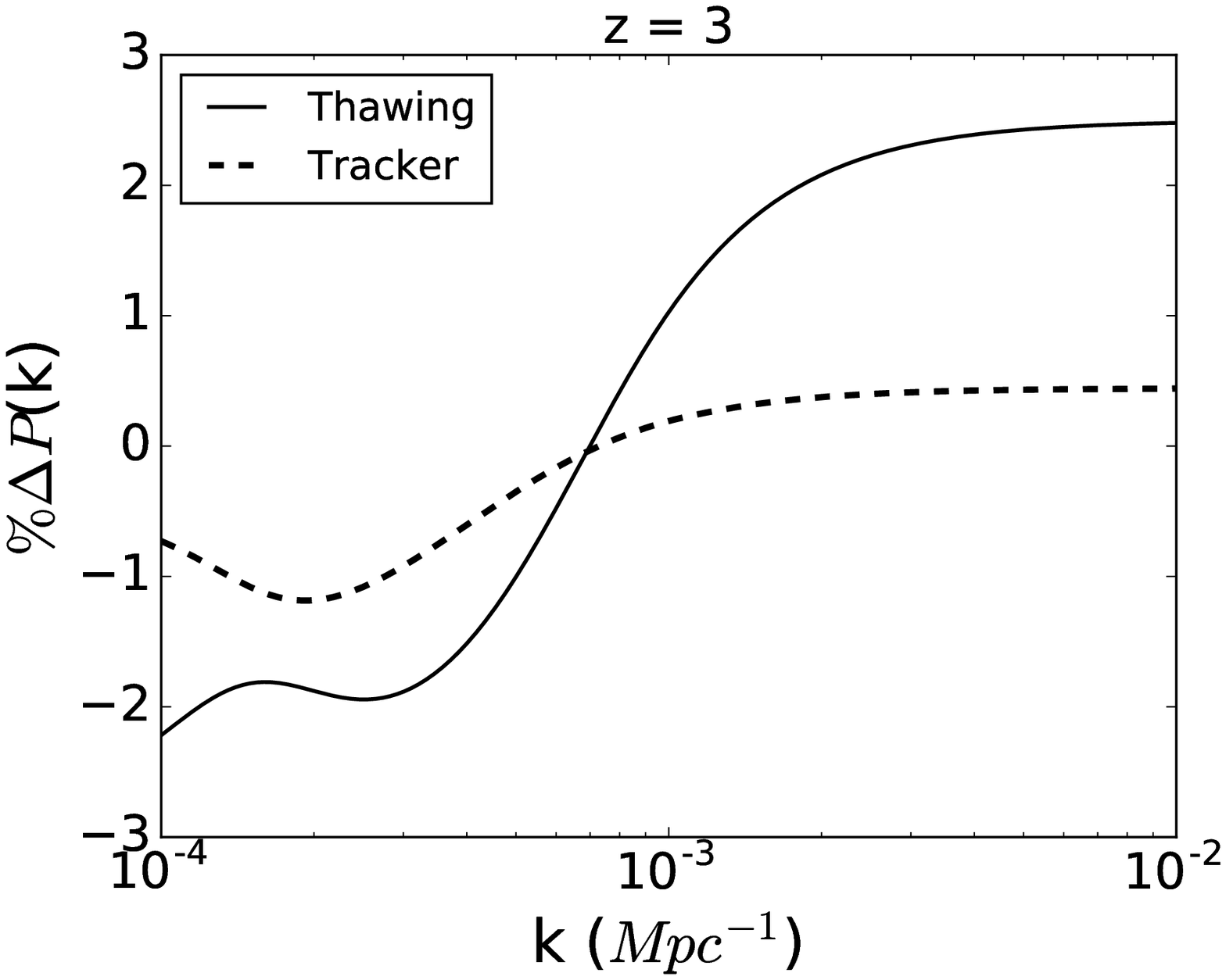,width=5.8 cm}
\end{tabular}
\caption{Percentage deviation in $ P (k) $ from $ \Lambda$CDM model as the function of redshifts both for thawing and tracker models respectively. Top most plots are standard matter power spectra $P_{s}$ given by eqn. (27), middle plots are for power spectra with Kaiser redshift space distortion term included and the bottom ones for full observed galaxy power spectra $P$ given by eqn (26) with GR corrections.
}
\end{figure*}
\end{center}

\noindent

In figure 9, we have shown percentage deviation of both thawing and tracker models in standard matter power spectrum, power spectrum with Kaiser term only and in the full relativistic observed galaxy power spectrum respectively from the $ \Lambda $CDM model taking all the cosmological parameters same as in section 5. As one can see, in thawer model, the deviation from $\Lambda$CDM is larger than corresponding deviation in tracking model on larger scales and for smaller redshifts. This is purely due to different GR corrections in these models. This difference between thawer and tracker model can be as large as $5-12\%$ depending upon the scales and redshifts. One smaller scales and for larger redshifts, this difference between thawer and tracker model reduces substantially and is mostly due to the difference in background evolution in these two models.

We should stress that these plots are obtained using specific choices of the parameters in both thawer and tracker models. The numbers can vary depending upon the values of the parameters. In an ideal scenario, one should first constrain these parameters in both the models from background observations and then use those constrained parameter space to see the difference between thawer and tracker models for the observed galaxy power spectrum. But this is beyond the scope of this paper and will be addressed in a separate paper.

\section{Conclusion}

Future survey like LSST, SKA etc have the potential to probe our universe on very large scales and also at very higher redshifts. This will give us the opportunity to probe the general relativistic effects on large scale structure formations. As we start probing universe on very large scales, we can not ignore the dark energy perturbations and hence all the GR corrections in observed galaxy power spectrum contain the contribution from dark energy perturbations. This is indeed very promising as it will be possible to distinguish $\Lambda$CDM model (where there is no dark energy perturbation) from other evolving dark energy models (where dark energy perturbations are present) in a completely new way  with new generation future surveys. Hence it is important to study the observed galaxy power spectra with relevant GR corrections for different non-$\Lambda$CDM dark energy models.

For tracking/freezing models,  this has done earlier by \citep{Duniya:2013eta}. In this work we extend this for thawing scalar field models for dark energy which is also a natural alternative to $\Lambda$CDM model. Interestingly these models can also give rise to transient acceleration. 

We set up a very general autonomous system of equations involving both the background as well as the perturbed universe. This is valid for any form of the potential irrespective of whether it is thawer model or tracker model. This set of equations are first of its kind and can be easily generalized to include other forms of matter like radiation, neutrinos etc. Subsequently we solve this system of equations with thawing type initial conditions for the scalar field evolution for various forms of scalar field potential. Our main aim is to see the effect of thawing scalar fields on observed galaxy power spectrum on large scales with different GR corrections and compare it to the $\Lambda$CDM model.

The gravitational potential  in scalar field model is enhanced from  $\Lambda$CDM value on large scales due to extra contribution from dark energy perturbation as determined by the Poisson equation. This extra contribution from dark energy perturbation is not present on small scales. Hence the small scale deviation from $\Lambda$CDM in scalar field model is always driven by difference in background expansion.  Due to the interplay of these two effects, in $P_{s}$ and $P_{ks}$, there is always suppression of power at large scales and enhancement of power on small scales in scalar field models in comparison to $\Lambda$CDM.

Once we add the GR correction term in the observed galaxy power spectra, the small scale behaviour remains the same ( which comes only due to background expansion), but on large scales, the suppression of power is increased by $9-10\%$ due to extra effect from GR correction, specifically from the term $\mathcal{A}$ which involves both the peculiar velocity as well as the gravitational potential. This deviation is expected to be probed by upcoming experiments like SKA.

We also compare the observed galaxy power spectra for thawing and tracking model on large scales assuming two specific potentials. We show that on large scales and for smaller redshifts,  thawer model can have larger suppression of power from $\Lambda$CDM than the tracker models. This shows that the GR corrections in these two models can be substantially different. On smaller scales and for larger redshifts, where the effect from background expansion dominates, the difference between these two models is not substantial. But we should stress that this difference between thawing and tracking models depends on the choice of the parameters in the potentials and unless we have specific constraints on these parameters, it is difficult to say anything conclusively.

\section{Acknowledgements}
B.R.D. thanks CSIR, Govt. of India for financial support through SRF scheme (No:09/466(0157)/2012-EMR-I). A.A.S acknowledges the support provided by Abdus Salam International Center For Theoretical Physics, Trieste, Italy through its Associate Program, where part of the work has been done. The authors also thank Sumit Kumar for valuable inputs regarding the python programming for the computational part of the problem. We also acknowledge the usage of HOPE-A Python Just-In-Time compiler for astrophysical computations \citep{2015A&C....10....1A} for our computation.

\bibliography{ref}

\end{document}